\documentclass[12pt]{iopart}
\usepackage{graphicx}
\usepackage{setspace}
\usepackage{lmodern}
\usepackage[english]{babel}
\usepackage[utf8]{inputenc}	
\usepackage{booktabs}
\usepackage{float}
\usepackage{graphicx}
\usepackage{eufrak}
\usepackage{stmaryrd}
\usepackage{amsfonts,amssymb}
\usepackage{multirow}
\usepackage{siunitx}
\usepackage{dsfont}
\usepackage{pgfplots}
\usepackage{wrapfig}
\usepackage[singlelinecheck=off]{subcaption}
\usepackage{caption}	
\usepackage{bm}
\usepackage[numbers]{natbib}
\usepackage{hyperref}
\usepackage{xcolor}
\usepackage{ulem}

\newcommand{\ket}[1]{\vert#1\rangle}
\newcommand{\bra}[1]{\langle #1\vert}
\newcommand{\abs}[1]{\left| #1\right|}
\usepackage{iopams}  
\def\aligned{\vcenter\bgroup\let\\\cr
\halign\bgroup&\hfil${}##{}$&${}##{}$\hfil\cr}
\def\endaligned{\crcr\egroup\egroup}

\begin{document}

\title[Two-qubit master equation: local vs global]{Local vs global master 
equation with common and separate baths: superiority of the global approach 
in partial secular approximation}

\author{Marco Cattaneo$^{1,2}$, Gian Luca Giorgi$^1$, Sabrina Maniscalco$^{2,3}$ 
and Roberta Zambrini$^1$}

\address{$^1$ Instituto de Física Interdisciplinar y Sistemas Complejos IFISC 
(CSIC-UIB), Campus Universitat Illes Balears, E-07122 Palma de Mallorca, Spain}
\address{$^2$ QTF Centre of Excellence, Turku Centre for Quantum Physics, 
Department of Physics and Astronomy, University of Turku, FI-20014 Turun 
Yliopisto, Finland}
\address{$^3$ QTF Centre of Excellence, Department of Applied Physics, School of 
Science, Aalto University, FI-00076 Aalto, Finland}
\ead{marcocattaneo@ifisc.uib-csic.es}
\vspace{10pt}
\begin{indented}
\item[]November 2019
\end{indented}

\begin{abstract}
Open systems of coupled qubits are ubiquitous in quantum physics. 
Finding a suitable master equation to describe their dynamics is therefore 
a crucial task that must be addressed with utmost attention. In the recent past, 
many efforts have been made toward the possibility of employing local master 
equations, which compute the interaction with the environment neglecting the 
direct coupling between the qubits, and for this reason may be easier to solve. 
Here, we provide a detailed derivation of the Markovian master equation for two 
coupled qubits interacting with common and separate baths, considering pure 
dephasing as well as dissipation. Then, we explore 
the differences between the local and global
master equation, showing that they intrinsically depend on the way we apply 
the 
secular approximation. Our results prove that the global approach with 
partial 
secular approximation always provides the most accurate choice for the master 
equation when Born-Markov approximations hold, even for 
small inter-system coupling constants. Using different master equations we compute the 
stationary heat current between 
two separate baths, the entanglement dynamics generated by a common bath, and 
the emergence of spontaneous synchronization, showing the importance of the accurate choice of approach.

\end{abstract}

%
%
%
%
%

\maketitle
\section{Introduction}
\label{sec:Intro}

Open quantum systems of two coupled qubits are of fundamental importance in many disparate fields, being for instance at the basis of the realization of multi-qubit 
gates for quantum computation \cite{Storcz2003,PhysRevLett.89.147902,Haddadfarshi_2016},
distributed quantum sensing  and  metrology \cite{Reiter2017,PhysRevX.7.041009}, and entanglement generation \cite{PhysRevLett.89.277901,Benatti2003a,Horn_2018}. Such systems have been experimentally simulated in a variety of platforms, including trapped ions \cite{Barreiro2011,Lin2013}, superconducting qubits \cite{PhysRevLett.116.240503}, or cavity QED arrays \cite{PhysRevX.7.011016}. They are also useful in the context of  quantum thermodynamics as they possess the minimum ingredients to realize  thermal machines \cite{Linden2010,Bohr_Brask_2015,PhysRevE.99.042135}.  Furthermore, in spite of their simplicity, they  allow  for the observation of  fundamental effects such as Dicke superradiance \cite{Dicke1954} or spontaneous quantum synchronization \cite{Giorgi2013}.
The derivation of the master equation describing the evolution of the qubits, and the subsequent 
search for an easy path to solve it, is therefore of the greatest importance.

While partial results investigating specific cases are available in the 
literature 
\cite{Ficek2002,Scala2008,Li2009a,Campagnano2010,Orth2010,Scala2011,Santos2014}, 
a general description of the problem is still missing. In this paper we provide a comprehensive analysis based on a miscroscopic derivation in the case of two qubits, addressing: the presence of dissipative as well as dephasing baths, which can be common and/or separate, 
and considering a sufficiently general interaction between the qubits not limited to a  Hamiltonian in rotating wave approximation (RWA),
and also allowing for frequency detuning. As this 
is often the case for most of the applications of the two-qubit problem, we will 
consider memory-less reservoirs, that is to say, we will study a 
\textit{Markovian} master equation.

Our detailed  derivation allows us  to establish the validity of the 
so-called \textit{local} approach for the master equation in comparison 
with a \textit{global} one in a rather general setting. The global approach  arises naturally when deriving the master equation from a microscopic model considering the full system Hamiltonian, i.e. in presence of interactions between its subsystems (here the two qubits), while the local one follows from the approximation which neglects these interactions. Recently, the problem of characterizing the range of 
applicability of the local rather than global master equation  has received much interest 
\cite{Levy2014,Trushechkin2016,Gonzalez2017,Hofer2017,Mitchison2018}, mostly 
related to the consistency of this decription in quantum thermodynamics. It is our aim to show here that an accurate application of the \textit{secular approximation} in the global approach always leads to a correct Markovian master equation, independently of the value of the coupling constant between the subsystems. The deep interconnection between a 
correct application of the secular approximation and the local vs global issue 
is discussed starting from the first principles of the derivation of the master 
equation. Deviations from the most accurate (global partial secular) approximation are illustrated
by looking at the open system dynamics as well as the steady state. 
 Moreover, we observe how the steady state heat current, the entanglement dynamics and the 
 presence of quantum beats or quantum synchronization vary when using distinct master equations, 
 so as to corroborate the validity (or inaccuracy) of each approach according to physical 
 considerations.

The global approach within a \textit{partial} 
secular approximation is compared with
the full secular approximation discussing the failure of the latter, which depends on the 
spacing between the energy levels of the free system Hamiltonian. The difference 
between common and separate baths plays here a central role. In this regard we 
prove that, in addition to the value of 
the qubit-qubit coupling constant already investigated in several works, the ratio between it and the detuning of the qubit frequencies plays 
an important role. Our conclusions are summarized in 
Table~\ref{tab:tabella} presented in the concluding remarks, and their validity 
exceeds the scenario of two coupled qubits, since for instance it holds for the 
case of coupled harmonic oscillators. The general discussion remains valid for 
more complex systems, composed of more than two subsystems as well.

In order to provide a self-contained presentation to tackle the issues arising in the local vs global problem, we first 
of all recall the derivation of the master 
equation, and the condition for the validity of each approximation, in Sec.~\ref{sec:Derivation}. 
The local vs global problem is set into the literature context and discussed in Sec.~\ref{sec:localGlobal}, 
first in general terms, and then for the specific case of two coupled qubits,
showing some relevant comparisons.
Sec.~\ref{sec:physicalQ} is devoted to the discussion of 
examples where the choice of the proper master equation is relevant for an accurate description
of physical quantities, such as the steady state heat current, 
the entanglement dynamics and the presence of quantum beats and quantum synchronization.
Finally, in Sec.~\ref{sec:Conclusions} we discuss some concluding remarks
further summarizing our findings in Table~\ref{tab:tabella}.

\section{Deriving the master equation}
\label{sec:Derivation}
\begin{figure}[t!]
\centering
\includegraphics[scale=0.31]{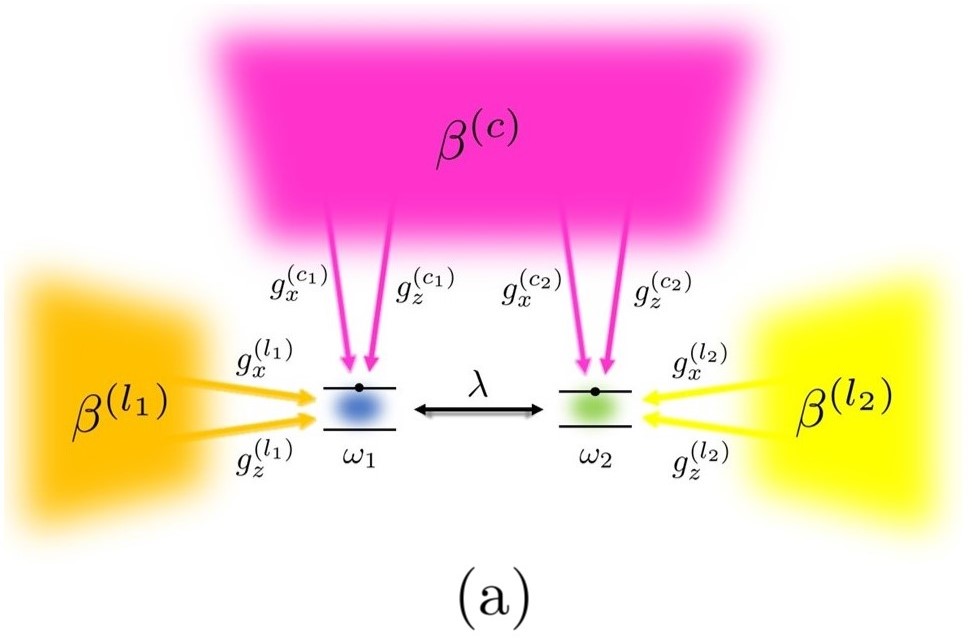}
\includegraphics[scale=0.23]{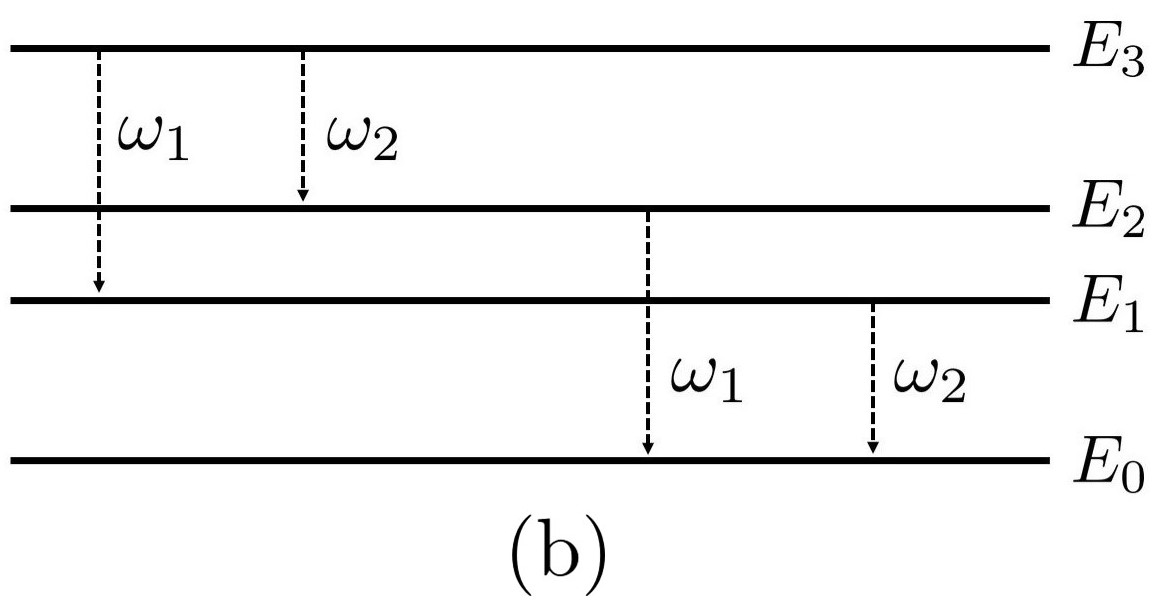}
\caption{(a): scheme of the two qubits interacting with thermal baths 
according to the Hamiltonian in equation~\ref{eqn:HamiltonianGeneral}, where the 
environments are characterized by the inverse temperatures $\beta^{(c)}$, 
$\beta^{(l_1)}$ and $\beta^{(l_2)}$, respectively for the common bath, the local 
bath on the first qubit and the local bath on the second qubit. An additional 
direct coupling between the qubits is mediated by the coupling constant 
$\lambda$. (b): diagram of the states of the system Hamiltonian 
equation~\ref{eqn:freeSystemHamInitRen}, setting $\lambda=0$ and with all the 
possible emission frequencies.}
\label{fig:schemeQubits}
\end{figure}

The aim of the present work is to address a general Markovian master equation 
for two qubits that can be detuned, exchange energy and are coupled to thermal 
baths: we consider both dephasing and dissipative interactions, and both 
separate and common baths, as in the pictorial representation in figure~\ref{fig:schemeQubits}a.

\subsection{Full Hamiltonian}
\label{sec:Hamiltonian}
Let us start by writing the free Hamiltonian of the system, in which we have set 
$\hbar=1$:
\begin{equation}
H'_S=\frac{\omega_1'}{2}\sigma_1^z+\frac{\omega_2'}{2}
\sigma_2^z+\lambda'\sigma_1^x\sigma_2^x,
\label{eqn:freeSystemHamInit}
\end{equation}
where $\omega_1'$ and $\omega_2'$ are the frequencies of respectively the first 
and second qubit, and $\lambda'$ is the qubit-qubit coupling constant.
We note that for the sake of generality we do not approximate the interaction by 
$\sigma_1^+\sigma_2^-+h.c.$, as in RWA. The 
generality of equation~\ref{eqn:freeSystemHamInit} is further discussed 
in~\ref{sec:generalitySystem}.

In order to work with dimensionless units, we renormalize the above Hamiltonian by 
$\omega_1'$, i.e. by the frequency of the first qubit:
\begin{equation}
\label{eqn:freeSystemHamInitRen}
H_S=\frac{\omega_1}{2}\sigma_1^z+\frac{\omega_2}{2}
\sigma_2^z+\lambda\sigma_1^x\sigma_2^x,~~~  \text{ with $\omega_1=1$, }
\end{equation}
$\omega_2=\omega_2'/\omega_1'$ and 
$\lambda=\lambda'/\omega_1'$. Through all the work, we assume that the 
renormalized qubit frequencies are of the same order, $\omega_2=O(1)$.

We now write the most general microscopic Hamiltonian of two coupled qubits interacting with 
common and separate thermal baths (consistently renormalized by the frequency of 
the first qubit):
\begin{equation}
\label{eqn:HamiltonianGeneral}
H=H_S+H_B^{(l_1)}+H_B^{(l_2)}+H_B^{(c)}+H_I,
\end{equation}
with
\begin{equation}
\label{eqn:interactingHgeneral}
\begin{aligned}
H_I=&\left(g_x^{(l_1)}\sigma_1^x+g_z^{(l_1)}\sigma_1^z\right)B^{(l_1)}
+\left(g_x^{(l_2)}\sigma_2^x+g_z^{(l_2)}\sigma_2^z\right)B^{(l_2)}\\
&+\left(g_x^{(c_1)}\sigma_1^x+g_x^{(c_2)}\sigma_2^x+g_z^{(c_1)}\sigma_1^z+g_z^{
(c_2)}\sigma_2^z\right)B^{(c)}.
\end{aligned}
\end{equation}
$H_B^{(l_1)}$ and $H_B^{(l_2)}$ refer to the free Hamiltonian of the local bath 
respectively on the first and on the second qubit, while $H_B^{(c)}$ is the free 
Hamiltonian of the common bath. They read:
\begin{equation}
\label{eqn:freeBathHam}
H_B^{(\alpha)}=\sum_k \Omega_{k,\alpha}a_{k,\alpha}^\dagger a_{k,\alpha},
\end{equation} 
where following the convention of quantum optics the summation over $k$ in the 
limit of infinite size bath represents as usual an integral over all the dense 
frequencies, and $\alpha=l_1,l_2,c$ indicates the specific bath. 

The bath operators appearing in the interaction Hamiltonian $H_I$ are given by
\begin{equation}
\label{eqn:bathOperators}
B^{(\alpha)}=\sum_k f_{k,\alpha}\left(a_{k,\alpha}^\dagger+a_{k,\alpha}\right),
\end{equation}
and the dissipative and dephasing couplings are mediated by the coefficients 
$g_x$ and $g_z$. For instance, $g_x^{(l_1)}$ is the dimensionless coupling 
constant describing the strength of the dissipative interaction between the 
first qubit and the respective local bath, and so on. For simplicity, we take 
the coupling constants real. Notice that we are using the standard denomination 
for ``dissipation'' and ``dephasing'', where the former refers to a coupling 
through $\sigma^x$, inducing both loss of energy and decoherence, while the 
latter denotes a coupling through $\sigma^z$, causing, at least in the uncoupled case, pure decoherence but no energy leak. 
Markovian master equations can be derived in the weak coupling limit of the 
qubit-bath interaction. Therefore, we introduce a constant $\mu$ such that each 
coupling strength appearing in the interaction Hamiltonian is at maximum of the 
order of $\mu$, i.e. $g_j^{(a)}=O(\mu)$ $\forall\, a=l_1,l_2,c_1,c_2$ and 
$j=x,z$, and we 
assume $\mu\ll 1$ (consistently in units of $\omega_1$).

The coupling coefficients $f_{k,\alpha}$ define the spectral density $J_\alpha(\omega)$ of each bath 
through:
\begin{equation}
\label{eqn:spectralDensity}
J_\alpha(\omega)=\sum_k f_{k,\alpha}^2\delta(\omega-\Omega_{k,\alpha}),
\end{equation} 
and we notice that the distinct dephasing and dissipative (and ``small'' 
$O(\mu)$) coupling $g_j^{(a)}$ are not included in equation~\ref{eqn:spectralDensity}.

One may wonder why, aiming at a complete description of any possible 
two-qubit system, we have considered the same bath inducing both dissipation and 
dephasing (in fact we could consider 6 instead of 3 baths). Assuming that 
different effects are due to different phenomena, a description employing a 
distinct bath for each of them should be necessary. Moreover, many uncorrelated 
environments may interact locally on each qubit, as it happens for example with 
a transmon qubit \cite{Koch2007a}, so why shall we describe them through a 
single bath, as done in equation~\ref{eqn:HamiltonianGeneral}? We anticipate that 
this assumption simplifies the notation and actually does not limit the 
following analysis, as we will be discussing in Sec.~\ref{sec:findingM}.

\subsection{Bloch-Redfield master equation in the secular approximation}
\label{sec:BlochRedfield}
In this section we will illustrate how to obtain a Markovian master equation 
starting from the microscopic Hamiltonian, stressing the validity of each 
approximation in order to get to a \textit{global} Bloch-Redfield master 
equation in the (partial) secular approximation. The possibility for a 
\textit{local} master equation will be discussed in Sec.~\ref{sec:localGlobal}.

Let us work in the interaction picture according to the free Hamiltonian 
$H_0=H_S+H_B$, where the full bath Hamiltonian is 
$H_B=H_B^{(l_1)}+H_B^{(l_2)}+H_B^{(c)}$. The Von-Neumann equation thus reads
\begin{equation}
\label{eqn:vonNeumann}
\frac{d}{dt}\rho(t)=-i[H_I(t),\rho(t)],
\end{equation}
where $\rho(t)$ and $H_I(t)$ denote the overall density matrix and the 
interaction Hamiltonian in the interaction picture representation (see 
\cite{BreuerPetruccione} for details).
By integrating equation~\ref{eqn:vonNeumann}, inserting it once again in 
equation~\ref{eqn:vonNeumann} and taking the partial trace as usual, we obtain an 
integro-differential equation for the reduced density matrix of the system 
$\rho_S(t)=\Tr_B[\rho(t)]$:
\begin{equation}
\label{eqn:integroDiffEq}
\frac{d}{dt}\rho_S(t)=-\int_0^t dt' \,\Tr_B[H_I(t),[H_I(t'),\rho(t')]],
\end{equation}
where $[H_I(t),\rho(0)]=0$, if the environment is in a thermal state. 

We now set the validity of essential approximations in order to get to a 
Markovian master equation:
	
\paragraph{Born approximation--} The interaction between system and environment is 
so weak that the state of the latter is \textit{almost} not perturbed by the 
coupling with the system. If the initial state of the overall system is the 
product state $\rho(0)=\rho_S(0)\otimes\rho_B$, the evolved state at a certain 
time $t$ is assumed product as well:
\begin{equation}
\label{eqn:BornApp}
\rho(t)\approx\rho_S(t)\otimes\rho_B.
\end{equation}
The approximation~\ref{eqn:BornApp} can be considered as a 
heuristic and intuitive way to obtain an important result, mathematically proven 
through the method developed by Nakajima \cite{Nakajima1958} and Zwanzig 
\cite{Zwanzig1960}. Indeed, it can be shown \cite{Rivas2010a} that, by inserting 
equation~\ref{eqn:BornApp} in equation~\ref{eqn:integroDiffEq}, we are neglecting terms of 
the order of $O(\mu^3)$, where $\mu$ is the coupling constant defined in the 
previous section. Therefore,
\begin{equation}
\label{eqn:integroDiffEqBis}
\frac{d}{dt}\rho_S(t)=-\int_0^t dt' 
\,\Tr_B[H_I(t),[H_I(t'),\rho_S(t')\otimes\rho_B]]+O(\mu^3).
\end{equation}
We point out that, while the full state $\rho(t)$ is not expected to remain factorized 
(as in equation~\ref{eqn:BornApp})
for long times \cite{Rivas2010a}, equation~\ref{eqn:integroDiffEqBis} is an exact result holding for 
any time $t$. 

Let us now decompose the interaction Hamiltonian in the interaction picture in 
the following way:
\begin{equation}
\label{eqn:decompositionIntH}
H_I(t)=\sum_\beta A_\beta(t)\otimes B_\beta(t),
\end{equation}
where $A_\beta(t)$ are system operators, while $B_\beta(t)$ are bath 
operators\footnote{The bath operators $B_\beta$ in 
equation~\ref{eqn:decompositionIntH} should not be confused with $B^{(\alpha)}$ defined in 
equation~\ref{eqn:interactingHgeneral} and \ref{eqn:bathOperators}: each $B_\beta$ is given by the product of the 
corresponding coupling constant $g_k^{(\alpha)}$ and the operator $B^{(\alpha)}$.}. If we make the change of variable $\tau=t-t'$ 
and insert equation~\ref{eqn:decompositionIntH} in equation~\ref{eqn:integroDiffEqBis}, 
after some algebra we obtain:
\begin{eqnarray}
\label{eqn:integroDiffEqTris}
\frac{d}{dt}\rho_S(t)=&-\sum_{\beta,\beta'}\int_0^t d\tau 
\left(\mathcal{B}_{\beta\beta'}(\tau)[A_\beta(t),A_{\beta'}
(t-\tau)\rho_S(t-\tau)]+h.c.\right) \\ 
&+O(\mu^3),\nonumber
\end{eqnarray}
having introduced the bath correlation function 
$\mathcal{B}_{\beta\beta'}(\tau)=\langle 
B_\beta(\tau)B_{\beta'}(0)\rangle_B=\Tr[B_\beta(\tau)B_{\beta'}(0)\rho_B]$, with 
the assumption that the bath is stationary, i.e. $[\rho_B,H_B]=0$. We are now 
ready to perform the next fundamental approximation.

\paragraph{Markov approximation--} We assume that the bath operators have a very 
short correlation time, and the correlation functions decay as 
$\abs{B_{\beta\beta'}(\tau)}\sim e^{-\tau/\tau_B}$. Remembering 
the weak coupling limit we then set $\tau_B\ll \tau_R$, i.e. the system will 
relax slowly with respect to the bath correlation functions, being $\tau_R$ the 
timescale over which the state in the interaction picture changes appreciably. 
Considering the highest order appearing in equation~\ref{eqn:integroDiffEqBis}, it is 
usually heuristically set
\begin{equation}
\label{eqn:timescaleSystem}
\tau_R=O(\mu^{-2}),
\end{equation}
where we remind that $\mu$ is the qubit-bath coupling constant renormalized by 
the frequency of the first qubit. The validity of the assumption needs often to 
be checked. For instance, in the limit of very high temperatures this might not 
be fulfilled, since a huge number of excitations would be available to interact 
with the system, making the decay rates very high as well. Nonetheless, there may exist the case in which, in the limit for the temperature $T\rightarrow \infty$, the autocorrelation functions of the bath decay faster than the relaxation time $\tau_R$, and therefore the Markov approximation is still valid. In this scenario, for $T\rightarrow\infty$ the autocorrelation functions of the bath are proportional to a Dirac delta, $\mathcal{B}_{\beta\beta'}(\tau)\propto\delta(\tau)$, and we recover the so-called singular-coupling limit \cite{BreuerPetruccione}.

If now we calculate the integral in equation~\ref{eqn:integroDiffEqTris} for a 
sufficiently large time $t^*\gg\tau_B$, such that $t^*$ is still way smaller 
than the time $\tau_R$ at which the state of the system in interaction picture 
changes appreciably, then we can safely replace $\rho_S(t-\tau)$ with 
$\rho_S(t)$ in the same equation, since the dynamics of $\rho_S(t)$ is way 
slower than the decay of $\mathcal{B}_{\beta\beta'}(\tau)$. For the same reason, we 
can extent the integral till infinity, since the added part will give a 
negligible contribution. This is the Markov approximation, which sets a 
resolution on the timescale of the dynamics for $t^*$, such that
\begin{equation}
\label{eqn:markovCondition}
\tau_B\ll t^*\ll\tau_R=O(\mu^{-2}).
\end{equation} 
This corresponds to defining a certain coarse-grained timescale of the 
evolution; indeed, the Markovian master equation can alternatively be derived by 
making averages on these coarse-grained time intervals, as recently discussed in 
\cite{Cresser2017a,Farina2019}.

Finally we write:
\begin{eqnarray}
\label{eqn:integroDiffEqQuater}
\frac{d}{dt}\rho_S(t)=&-\sum_{\beta,\beta'}\int_0^\infty d\tau 
\left(\mathcal{B}_{\beta\beta'}(\tau)[A_\beta(t),A_{\beta'}(t-\tau)\rho_S(t)]
+h.c.\right)\\
&+o(\mu^2).\nonumber
\end{eqnarray}
Unfortunately, to the best of our knowledge a precise order for the remainder 
neglected in the Markov approximation equation~\ref{eqn:integroDiffEqQuater} has not 
been reported in general. 
An interesting bound is however provided in a recent 
paper \cite{Albash2012}, where instead of equation~\ref{eqn:markovCondition} the 
authors consider the tighter condition $\tau_B\ll\mu^{-1}$. In 
general, assuming the condition in equation~\ref{eqn:markovCondition}, the approximated master equation neglects a remainder 
of order higher than $O(\mu^2)$ (that from now on, we will drop); 
a special care in checking the validity of the Markov approximation in each 
specific case is anyway indispensable.

We will now further decompose the interaction Hamiltonian 
equation~\ref{eqn:decompositionIntH} by introducing the \textit{jump operators} 
associated to each system operator $A_\beta$:
\begin{equation}
\label{eqn:jumpOp}
A_\beta(\omega)=\sum_{\epsilon'-\epsilon=\omega}\ket{\epsilon}\bra{\epsilon}{
A_\beta}\ket{\epsilon'}\bra{\epsilon'},
\end{equation}
where $\{\ket{\epsilon}\}_\epsilon$ is the basis of the eigenvectors of the 
system Hamiltonian $H_S$. The following properties hold:
\begin{equation}
\label{eqn:propJumpOp}
A_\beta^\dagger(\omega)=A_\beta(-\omega),\qquad \sum_\omega 
A_\beta(\omega)=\sum_\omega A_\beta^\dagger(\omega)=A_\beta.
\end{equation}

By writing equation~\ref{eqn:integroDiffEqQuater} with the time-evolved jump 
operators, we get to the \textit{Bloch-Redfield master equation}
\begin{eqnarray}
\label{eqn:preSecular}
\frac{d}{dt}\rho_S(t)=\sum_{\omega,\omega'}\sum_{\beta,\beta'}e^{
i(\omega'-\omega)t}\Gamma_{\beta\beta'}(\omega)&\left(A_{\beta'}
(\omega)\rho_S(t)A_\beta^\dagger(\omega')\right.\\
&\left.-A_\beta^\dagger(\omega')A_{\beta'}(\omega)\rho_S(t)\right)+h.c.\,,
\nonumber
\end{eqnarray}
where we have introduced the one-side Fourier transform of the bath correlation 
functions,
\begin{equation}
\label{eqn:oneSideBath}
\Gamma_{\beta\beta'}(\omega)=\int_0^\infty dt' e^{i\omega 
t'}\mathcal{B}_{\beta\beta'}(t').
\end{equation}

\paragraph{Secular approximation--} The evolution of the state of the system 
$\rho_S(t)$ has, in the interaction picture, a typical relaxation timescale of 
the order of the square of the inverse of the coupling strength $\mu$, as stated 
in equation~\ref{eqn:timescaleSystem}. If there exist values of $\omega'$ and 
$\omega$ in equation~\ref{eqn:preSecular} being coarse-grained in time as from 
equation~\ref{eqn:markovCondition}, i.e.
\begin{equation}
\label{eqn:secularAppCond}
\exists\,t^*\textnormal{ such that }\abs{\omega'-\omega}^{-1}\ll t^*\ll 
\tau_R=O(\mu^{-2}),
\end{equation}
then the terms in equation~\ref{eqn:preSecular} oscillating with frequency 
$\omega'-\omega$ will not give any significant contribution to the system evolution, since 
by integrating equation~\ref{eqn:preSecular} for a time $t^*$ such that 
$\abs{\omega'-\omega}^{-1}\ll t^*\ll \tau_R$ the fast-oscillating quantities 
vanish. Equation~\ref{eqn:secularAppCond} corresponds to a refinement of the coarse-grain condition written in equation~\ref{eqn:markovCondition}. Indeed, a slightly different approach to the derivation of the master equation makes use of a unique coarse-grained average, including both the Markov and the secular approximation (see for instance Refs.~\cite{Cresser2017a,Farina2019}).
Notice that the interaction picture is particularly suited to distinguish the terms bringing 
a negligible contribution to the evolution of the system.

Neglecting the fast oscillating terms in the interaction picture is usually 
referred to as \textit{secular approximation}. Unfortunately, it is easy to run 
into a nomenclature issue in the literature: in many works we can find the name 
``secular approximation'' for the removal of \textit{all} the terms in 
equation~\ref{eqn:preSecular} for which $\omega'\neq\omega$, without questioning the 
validity of equation~\ref{eqn:secularAppCond}. This is of course feasible for all the 
systems in which the relevant frequencies are well-spaced, i.e. 
$\abs{\omega'-\omega}\gg \tau_R^{-1}\approx\mu^2$ for any $\omega',\omega$, but it 
might lead to confusion in other cases, as we will discuss in 
Sec.~\ref{sec:localGlobal}. 

To avoid confusion, we will call \textit{full secular} the approximation for 
which we neglect all the terms in equation~\ref{eqn:preSecular} with 
$\omega'\neq\omega$, while we will employ the name \textit{partial secular 
approximation} for the cases in which we keep some slowly rotating terms with 
$\omega'\neq\omega$, for which the relation in equation~\ref{eqn:secularAppCond} would actually fail. 
For the sake of clarity, when discussing concrete examples throughout the paper 
we will name the master equation with cross terms often retained in the 
secular approximation~\ref{eqn:secularAppCond} as master equation in partial secular approximation
(also in regimes where these terms could be neglected).

After some algebra, the Bloch-Redfield master equation equation~\ref{eqn:preSecular} 
may be rewritten in the Schr\"odinger picture as
\begin{equation}
\label{eqn:BlochRedfieldSchrodinger}
\frac{d}{dt}\rho_S(t)=-i[H_S+H_{LS},\rho_S(t)]+D[\rho_S(t)],
\end{equation}
where we have introduced the \textit{Lamb-Shift Hamiltonian}:
\begin{equation}
\label{eqn:lambShift}
H_{LS}=\sum_{\omega,\omega'}\sum_{\beta,\beta'} 
S_{\beta\beta'}(\omega,\omega')A_{\beta}^\dagger(\omega')A_{\beta'}(\omega),
\end{equation}   
and the \textit{dissipator} of the master equation, responsible for the energy 
losses of the system:
\begin{equation}
\label{eqn:dissipator}
\mathcal{D}(\rho_S)=\sum_{\omega,\omega'}\sum_{\beta,\beta'} 
\gamma_{\beta\beta'}(\omega,\omega')\left(A_{\beta'}(\omega)\rho_S 
A_\beta^\dagger(\omega')-\frac{1}{2}\{A_\beta^\dagger(\omega') 
A_{\beta'}(\omega),\rho_S\}\right),
\end{equation}
with
\begin{equation}
\label{eqn:Coefficients}
\begin{aligned}
S_{\beta\beta'}(\omega,\omega')&=\frac{\Gamma_{\beta\beta'}(\omega)-\Gamma_{
\beta'\beta}^*(\omega')}{2i},\\
\gamma_{\beta\beta'}(\omega,\omega')&=\Gamma_{\beta\beta'}(\omega)+\Gamma_{
\beta'\beta}^*(\omega').
\end{aligned}
\end{equation}
Notice that, prior to the secular approximation, the Lamb-Shift Hamiltonian is 
not Hermitian and contains imaginary terms as well, and we do not have a 
``purely dissipative'' dissipator. 

By employing the full secular approximation and coming back to the Schr\"odinger 
picture, the Lamb-Shift Hamiltonian and dissipator read:
\begin{equation}
\label{eqn:masterEqFullSecular}
\begin{aligned}
H_{LS}&=\sum_{\omega}\sum_{\beta,\beta'} 
S_{\beta\beta'}(\omega,\omega)A_{\beta}^\dagger(\omega)A_{\beta'}(\omega),\\
\mathcal{D}(\rho_S)&=\sum_{\omega}\sum_{\beta,\beta'} 
\gamma_{\beta\beta'}(\omega,\omega)\left(A_{\beta'}(\omega)\rho_S 
A_\beta^\dagger(\omega)-\frac{1}{2}\{A_\beta^\dagger(\omega) 
A_{\beta'}(\omega),\rho_S\}\right).\\
\end{aligned}
\end{equation}

The master equation~\ref{eqn:BlochRedfieldSchrodinger} with Lamb-Shift 
Hamiltonian and dissipator given by equation~\ref{eqn:masterEqFullSecular} is written 
in the GKLS form \cite{Gorini1976a,Lindblad1976,Chruscinski2017a}, and it 
therefore generates a dynamical semigroup, i.e. a perfectly Markovian evolution. 
On the other hand, this is a strong condition which is not necessary to get a GKLS form of the master equation.
In fact, some very recent papers \cite{Cresser2017a,Farina2019} have shown that performing an 
accurate partial secular approximation leads to a GKLS master equation as well 
(see also \cite{Galve2017,bellomo2017}).
An interesting observation is that the partial secular approximation condition (\ref{eqn:secularAppCond}) 
is sufficient to remove fast oscillating terms leading to a dissipator (\ref{eqn:dissipator}) where
 terms $\left(A_{\beta'}(\omega)\rho_S 
A_\beta(\omega')-\frac{1}{2}\{A_\beta(\omega') 
A_{\beta'}(\omega),\rho_S\}\right)$, with $\omega$ and $\omega^\prime$ with 
the same sign (which would produce a squeezing-like effect), are prevented. In fact the fastest terms, more susceptible to fulfill the 
condition (\ref{eqn:secularAppCond}),  will oscillate at frequency $|\omega-(-\omega^\prime)| $
(again with $\omega$ and $\omega^\prime$ with the same sign): if this is the case, then all terms 
$A_{\beta'}(\omega)\rho_S 
A_\beta^\dagger(-\omega')\equiv A_{\beta'}(\omega)\rho_S 
A_\beta(\omega')$  will be consistently neglected.

Finally, let us term the master equation~\ref{eqn:preSecular} in partial 
secular approximation, which may be rewritten in the form of 
equation~\ref{eqn:BlochRedfieldSchrodinger}, ``global master equation with 
partial secular approximation''. We stress the fact that this master equation 
is derived within the Born-Markov approximations. The latter can be more delicate to 
assess
and  unphysical effects can arise as signatures of an inaccurate Markovian 
description of an intrinsically non-Markovian evolution \cite{Hartmann2019}.
On the other hand,
in general it is immediate to establish the validity of the condition  
(\ref{eqn:secularAppCond})
to get an equation in the partial secular approximation. Also the inaccurate 
secular approximation,
i.e. out of the validity region (\ref{eqn:secularAppCond}), can lead to 
unphysical effects, as can be displayed by
a full secular master equation
with respect to a partial secular one.



\subsection{Diagonalizing the system Hamiltonian and finding the jump operators}
\label{sec:findingM}
As discussed in the previous section, diagonalizing the system Hamiltonian $H_S$ is a necessary step to derive the Markovian master equation, 
since it allows us to write in the correct form the jump operators defined in 
equation~\ref{eqn:jumpOp} which describe the effects of the interaction with the 
baths.

\subsubsection{No direct coupling}
\label{sec:noDirect}
\begin{figure}
\centering
\includegraphics[scale=0.30]{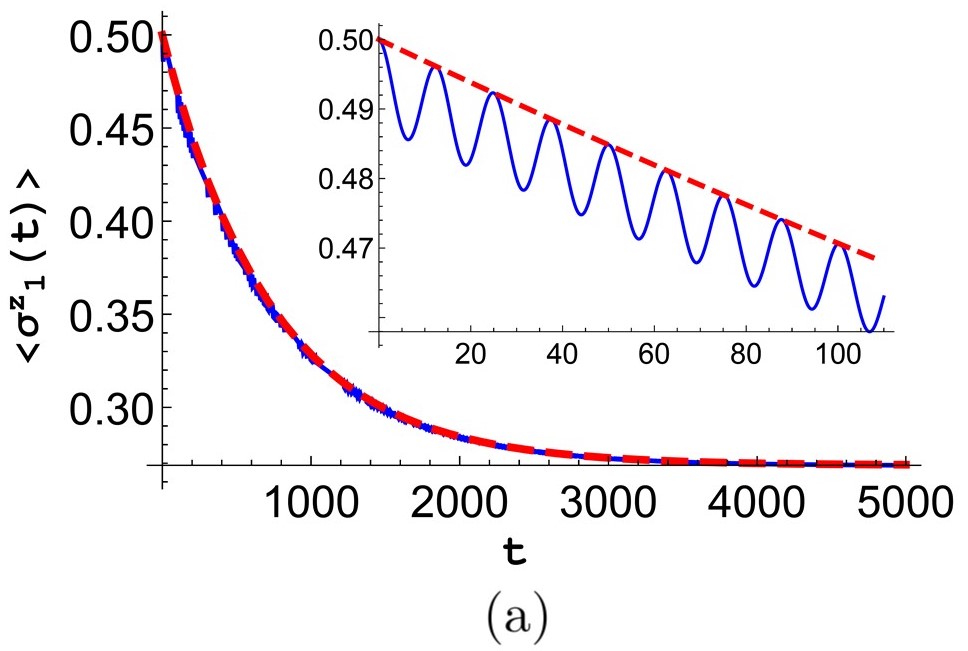}
\includegraphics[scale=0.30]{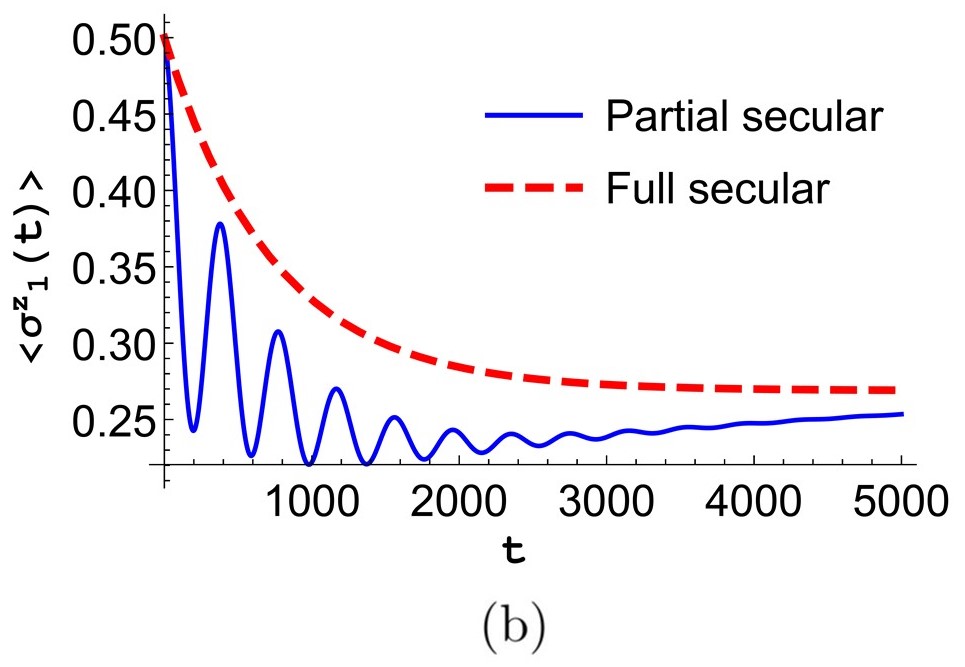}
\caption{ In the case of uncoupled qubits, i.e. $\lambda=0$, mean 
value of the population of the first qubit as a function of time, starting from 
the state $\rho_0=\rho_{OV}\otimes\rho_{OV}$, with 
$\rho_{OV}=1/2(\ket{0}\bra{0}+\ket{0}\bra{1}+\ket{1}\bra{0}+\ket{1}\bra{1})$, 
and the dissipative common bath is in a thermal state with $\beta^{(c)}=1$; all the 
other baths are switched off ($g_x^{(c_1)}=g_x^{(c_2)}=1$ and 
all the other coupling constants vanish). (a): case of big detuning with $\omega_1=1,$ $\omega_2=0.5$, $\omega_-=0.5$. (b): case 
of small detuning with $\omega_1=1,$ $\omega_2=0.99$, $\omega_-=0.01$. In figure~2a,  
the full secular approximation (dashed red) provides a correct way to describe 
the evolution, although the tiny oscillations given by the partial secular 
approximation in which we keep the cross terms (solid blue) can be observed by zooming to a proper small time 
interval (inset), which anyway cannot be resolved in the timescale defined by the coarse-graining. In figure~2(b), due to the small detuning, the full secular approximation (dashed red) 
fails and it leads to a completely different evolution with respect to the 
partial secular (solid blue).}
\label{fig:evNoDir}
\end{figure}
Let us start with the simplest case, i.e. in the absence of a direct coupling 
between the qubits: setting $\lambda=0$ in equation~\ref{eqn:freeSystemHamInitRen}, 
the system Hamiltonian reads:
\begin{equation}
\label{eqn:systemHamNoDir}
H_S=\frac{\omega_1}{2}\sigma_1^z+\frac{\omega_2}{2}\sigma_2^z,
\end{equation}
which is already diagonal in the ``canonical basis'' 
$\{\ket{00},\ket{01},\ket{10},\ket{11}\}$, with eigenvalues respectively 
$E_0=-\omega_+/2$, $E_1=-\omega_-/2$, $E_2=\omega_-/2$, $E_3=\omega_+/2$, where 
$\omega_\pm=\omega_1\pm\omega_2$ and (without losing generality) we set 
$\omega_1>\omega_2$. 

In the interaction Hamiltonian we can find the system operators $\sigma_j^x$ and 
$\sigma_j^z$ coupled to the bath operators, with $j=1,2$. Their decomposition in 
terms of jump operators is readily written according to equation~\ref{eqn:jumpOp}:
\begin{equation}
\label{eqn:jumpOperatorsNoDir}
\begin{aligned}
&\sigma_j^x(\omega_j)=\sigma_j^-,
\quad\sigma_j^x(-\omega_j)=\sigma_j^+\;\Rightarrow\;\sigma_j^x=\sigma_j^-+\sigma
_j^+,\\
&\sigma_j^z(0)=\sigma_j^z\;\Rightarrow\;\sigma_j^z=\sigma_j^z,\\
\end{aligned}
\end{equation}
with $j=1,2$. Equation~\ref{eqn:jumpOperatorsNoDir} describes the possible emission and absorption 
processes of the system, depicted in figure~\ref{fig:schemeQubits}b.

A self-consistent secular approximation depends on the detuning between the 
qubits. In the case in which there is a small detuning, such that 
$\omega_1-\omega_2$ is not way greater than $\mu^2$, we cannot employ the full 
secular approximation, but we need to rely on a partial secular approximation in 
which we keep in equation~\ref{eqn:preSecular} slow terms with 
$\omega'-\omega=\pm(\omega_1-\omega_2)$. The validity of the full secular 
approximation for big detuning and its breakdown in the opposite scenario are respectively  
shown in figures~\ref{fig:evNoDir}a and~\ref{fig:evNoDir}b. 
The most evident difference between partial and full secular descriptions
in the regime in which the latter fails (figure~\ref{fig:evNoDir}b) is the presence of 
so-called \textit{quantum beats}, i.e. of oscillations of the population of the excited 
state of the qubit. The quantum beats are a well-known phenomenon occurring 
during a superradiant emission \cite{gross1982superradiance} and predicted also 
for two non-identical atoms \cite{ficek1987quantum} as in our case. The master 
equation with partial secular approximation correctly describes them, while the 
full secular one is too crude and not able to reproduce the beats, 
leading to a completely smooth 
evolution. Further 
cases in wich  the full secular approximation is not suitable 
will be discussed in Sec.~\ref{sec:physicalQ}.

Starting from equation~\ref{eqn:preSecular} and employing the notation of 
equations~\ref{eqn:lambShift} and~\ref{eqn:dissipator}, the master equation finally 
reads:
\begin{eqnarray}
\label{eqn:masterEqNoDir}
\frac{d}{dt}\rho_S(t)=&-i[H_S+H_{LS},\rho_S(t)] \nonumber \\
&+\underbrace{\sum_{j,k=1,2}\gamma_{jk}\left(\sigma_j^-\rho_S(t)\sigma_k^+-\frac
{1}{2}\{\sigma_k^+\sigma_j^-,\rho_S(t)\}\right)}_{\textnormal{emission}}\\
&+\underbrace{\sum_{j,k=1,2}\tilde{\gamma}_{jk}
\left(\sigma_j^+\rho_S(t)\sigma_k^- 
-\frac{1}{2}\{\sigma_k^-\sigma_j^+,\rho_S(t)\}\right)}_{\textnormal{absorption}}
\nonumber\\
&+\underbrace{\sum_{j,k=1,2}\eta_{jk}
\left(\sigma_j^z\rho_S(t)\sigma_k^z-\frac{1}{2}\{\sigma_k^z\sigma_j^z,\rho_S(t)\}\right)}_{\textnormal{dephasing}},
\nonumber
\end{eqnarray}
where the Lamb-shift Hamiltonian is given by:
\begin{equation}
\label{eqn:lambShiftNoDir}
\begin{aligned}
H_{LS}=&\sum_{jk}\left(s_{jk}\sigma_k^+\sigma_j^-+\tilde{s}_{jk}
\sigma_k^-\sigma_j^+\right)+2s_0 \sigma_1^z\sigma_2^z\\
=&\frac{s_1}{2}\sigma_1^z+\frac{s_2}{2}
\sigma_2^z+s_+\sigma_1^+\sigma_2^-+s_-\sigma_1^-\sigma_2^++2s_0\sigma_1^z\sigma_
2^z,
\end{aligned}
\end{equation}
and the coefficients of the master equation are	 
presented in~\ref{sec:coeff}.

By looking at the jump operators in equation~\ref{eqn:jumpOperatorsNoDir} and at the 
partial secular approximation performed on equation~\ref{eqn:preSecular} we can now address the claim in Sec.~\ref{sec:Hamiltonian} about the simplifying 
choice of considering one single bath inducing both dephasing and dissipation. 
The point is that considering two distinct baths rather than a single one  
 is in general not needed (unless dephasing and dissipation need 
to be considered with different spectral density or baths temperatures), 
and would lengthen all expressions. More in detail, considering multiple equivalent and independent baths could
at most affect
 the coefficients $\Gamma_{\beta\beta'}(\omega)$ (see~\ref{sec:coeff} 
for their specific form): then we would 
have $\Gamma_{\beta\beta'}(\omega)=0$ for any system operators $A_\beta$ and 
$A_{\beta'}$ coupled to distinct baths through $B_\beta$ and $B_{\beta'}$, since 
there are no correlations between the two baths. On the contrary, if $B_\beta$ 
and $B_{\beta'}$ are operators of the same bath, then 
$\Gamma_{\beta\beta'}(\omega)$ does not vanish a priori and there could be a 
case in which $A_\beta\neq 
A_{\beta'}$ but $\Gamma_{\beta\beta'}(\omega)\neq 0$. Let us for instance consider the 
coupling with the local bath on the first qubits: 
$H_I^{(l_1)}=\left(g_x^{(l_1)}\sigma_1^x+g_z^{(l_1)}\sigma_1^z\right)B^{(l_1)}$.
 What if the dissipation would be induced by a bath different than the 
dephasing one? Looking at equation~\ref{eqn:jumpOperatorsNoDir} we can see that the 
operators $\sigma_1^x$ and $\sigma_1^z$ may in theory couple in 
equation~\ref{eqn:preSecular} with a non-zero coefficient 
$\Gamma_{xz}^{(l_1)}(\omega_1)$ or $\Gamma_{xz}^{(l_1)}(0)$, but their 
corresponding terms would vanish because of the partial secular approximation, 
since $\abs{\omega'-\omega}=\abs{\pm \omega_1}\gg 1/\tau_R$. It is easy to 
recognize that this argument holds for any case where dephasing and dissipation 
could arise from different baths. Therefore, for equivalent but independent baths, the simplified 
Hamiltonian equation~\ref{eqn:HamiltonianGeneral} can be assumed. Otherwise, considering 6 baths (instead of 3) would lead to different values of the bath correlation functions 
in equation~\ref{eqn:coeffNoDir}, but not change the structure of the master 
equation\footnote{Namely, the values of the $\Gamma$ for $\eta_{jk}$ and $s_0$ would be 
different from the $\Gamma$ associated to all the other coefficients. As a 
reference for the actual computation of the coefficients given a bath at inverse 
temperature $\beta$, see \cite{BreuerPetruccione}.}. We will reach the same 
conclusion in presence of qubits coupling, 
apart from a singular case (corresponding to a very specific parameter choice, when the condition 
in equation~\ref{eqn:conditionCrossing} holds).

Following the same path, we can readily see that if different sources, 
associated to different baths, would induce, let us say, independent dissipations 
on the same qubit, by assuming a single dissipative local bath we are not losing 
generality, since the effects of the multiple baths would not change the form of 
the master equation, but at most the value of the coefficients: the effects of 
independent baths would just sum, i.e. the final decay rate would be the sum of 
the decay rates given by each single bath.
The argument we have just discussed is reflected in the values of the 
coefficients in equation~\ref{eqn:coeffNoDir}.

\subsubsection{Direct coupling}
\label{sec:directC}
The case in which we have a direct qubit-qubit coupling should not in principle 
be more complex, since all we need to do is to diagonalize a $4\times 4$ matrix, 
find the corresponding eigenvalues and eigenvectors and work in the new basis. 
The system Hamiltonian $H_S$ now reads:
\begin{equation}
\label{eqn:systemHamDir}
H_S=\frac{\omega_1}{2}\sigma_1^z+\frac{\omega_2}{2}\sigma_2^z+\lambda 
\sigma_1^x\sigma_2^x,
\end{equation}
and the corresponding matrix in the canonical basis 
$\{\ket{11},\ket{10},\ket{01},\ket{00}\}$ is written as
\begin{equation}
\label{eqn:matrixHcan}
H_S=\left(\begin{array}{cccc}
\omega_+/2&0&0&\lambda\\
0&\omega_-/2&\lambda&0\\
0&\lambda&-\omega_-/2&0\\
\lambda&0&0&-\omega_+/2
\end{array}\right),
\end{equation}
with $\omega_\pm=\omega_1\pm\omega_2$.

We can easily diagonalize equation~\ref{eqn:matrixHcan} by finding the eigenvalues:
\begin{equation}
\label{eqn:eigenValDir}
\begin{aligned}
E_0&=-\sqrt{\lambda^2+\omega_+^2/4},\qquad\qquad
E_1&=-\sqrt{\lambda^2+\omega_-^2/4},\\
E_2&=+\sqrt{\lambda^2+\omega_-^2/4},\qquad\qquad
E_3&=+\sqrt{\lambda^2+\omega_+^2/4},\\
\end{aligned}
\end{equation}
with associated eigenvectors \cite{Scala2008} 
\begin{equation}
\label{eqn:eigenVecDir}
\begin{aligned}
\ket{e_0}&=-\sin\theta\ket{11}+\cos\theta\ket{00},\qquad&
\ket{e_1}&=-\sin\phi\ket{10}+\cos\phi\ket{01},\\
\ket{e_2}&=+\cos\phi\ket{10}+\sin\phi\ket{01},\qquad&
\ket{e_3}&=+\cos\theta\ket{11}+\sin\theta\ket{00}.
\end{aligned}
\end{equation}
where the parameters $\theta$ and $\phi$ are given by
\begin{equation}
\label{eqn:paramTheta}
\begin{aligned}
&\sin{2\theta}=\frac{\lambda}{E_3},\qquad 
\cos{2\theta}=\frac{\omega_+/2}{E_3},\\
&\sin{2\phi}=\frac{\lambda}{E_2},\qquad \cos{2\phi}=\frac{\omega_-/2}{E_2}.\\
\end{aligned}
\end{equation}

Once we know the spectral decomposition of the Hamiltonian, we can proceed to 
calculate the jump operators associated with each system operator appearing in 
the interaction Hamiltonian, i.e. $\sigma_j^x$ and $\sigma_j^z$ with $j=1,2$. 
The explicit form of each jump operator is given in~\ref{sec:jump}. With the aim 
at a complete description of the problem, we also consider the possibility that 
two of the eigenstates of the system are almost degenerate, which would happen 
if $\omega_-\ll 1$ and $\lambda\ll 1$, as it can be seen from 
equation~\ref{eqn:eigenValDir}. In this case, some additional terms beyond the full 
secular approximation need to be consistently kept, as fully observed 
in~\ref{sec:jump}.

With these prescriptions, the master equation reads:
\begin{eqnarray}
\label{eqn:masterEqDir}
\frac{d}{dt}\rho_S(t)=&-i[H_S+H_{LS},\rho_S(t)]\nonumber \\
&+\sum_{j,k=I,II \atop 
m,n=1,2}\gamma_{jk}^{mn}
\Big(\sigma_m^x(\omega_j)\rho_S(t)\sigma_n^x(-\omega_k)\nonumber\\
&\qquad\qquad\qquad-\frac{1}{2}\{\sigma_n^x(-\omega_k)\sigma_m^x(\omega_j),
\rho_S(t)\}\Big)\nonumber	\\
&+\sum_{j,k=I,II \atop 
m,n=1,2}\tilde{\gamma}_{jk}^{mn}
\Big(\sigma_m^x(-\omega_j)\rho_S(t)\sigma_n^x(\omega_k)\\
&\qquad\qquad\qquad-\frac{1}{2}\{\sigma_n^x(\omega_k)\sigma_m^x(-\omega_j),
\rho_S(t)\}\Big)\nonumber\\
&+\sum_{j,k=0,\pm{IV} \atop 
m,n=1,2}\eta_{jk}^{mn}
\Big(\sigma_m^z(\omega_j)\rho_S(t)\sigma_n^z(-\omega_k)\nonumber\\
&\qquad\qquad\qquad-\frac{1}{2}\{\sigma_n^z(-\omega_k)\sigma_m^z(\omega_j),
\rho_S(t)\}\Big)\nonumber\\
&+\sum_{j=\pm{III} \atop 
m,n=1,2}\zeta_{j}^{mn}
\Big(\sigma_m^z(\omega_j)\rho_S(t)\sigma_n^z(-\omega_j)\nonumber\\
&\qquad\qquad\qquad-\frac{1}{2}\{\sigma_n^z(-\omega_j)\sigma_m^z(\omega_j),
\rho_S(t)\}\Big),\nonumber	
\end{eqnarray}
where the jump operators and relative frequencies $\sigma_m^x(\omega_j)$ are 
defined in equations~\ref{eqn:jumpFrequencies} and~\ref{eqn:jumpOpDir}, and we are 
using the short notation $\omega_{-IV}=-\omega_{IV}$, 
$\omega_{-III}=-\omega_{III}$ and $\omega_0=0$. The Lamb-Shift Hamiltonian is 
given by:
\begin{eqnarray}
\label{eqn:LambShiftDir}
H_{LS}=&\sum_{j,k=I,II \atop m,n=1,2} 
\left(s_{jk}^{mn}\sigma_n^x(-\omega_k)\sigma_m^x(\omega_j)+\tilde{s}_{jk}^{mn}
\sigma_n^x(\omega_k)\sigma_m^x(-\omega_j)\right)\nonumber\\
&+\sum_{j,k=0,\pm{IV} \atop m,n=1,2} 
r_{jk}^{mn}\sigma_n^z(-\omega_k)\sigma_m^z(\omega_j)\\
&+\sum_{j=\pm{III} \atop m,n=1,2} 
u_j^{mn}\sigma_n^z(-\omega_j)\sigma_m^z(\omega_j).\nonumber
\end{eqnarray}
The coefficients of the master equation are listed in 
equation~\ref{eqn:coefficientsMasterEqDir} in~\ref{sec:coeff}.

\section{Local vs global: an in-depth discussion}
\label{sec:localGlobal}
A debate about the validity of the \textit{local} rather than the 
\textit{global} description of an open quantum system has arisen since the early 
era of the field: to the best of our knowledge, the first discussions about how to derive a global master equation accounting for the inter-system interactions date back to the early seventies \cite{walls1970higher,schwendimann1972interference,Carmichael_1973}. Twenty years later, Cresser observed the 
failure of the local approach to describe a lossy Jaynes-Cummings model~\cite{Cresser1992}, terming ``phenomenological master equation'' what it is nowadays 
usually called ``local master equation''. Quite the same issue has been 
addressed in some more recent papers \cite{Scala2007,Scala2007a}, extending the 
analysis to three coupled Josephson junctions \cite{Migliore2011} or coupled 
harmonic oscillators \cite{Rivas2010a}, while the 3-level atom has been 
investigated in Ref.~\cite{Zoubi2003}. Some comments about the validity of the 
local approach to describe energy transport in chains of harmonic oscillators or 
spins appear in Refs.~\cite{Saito2000,Henrich2005a,Wichterich2007}.

In the past few years, a renewed interest in the topic has grown, also because 
of a paper in 2014 suggesting that the local approach was breaking the second 
law of thermodynamics in a thermal machine composed of two quantum nodes
\cite{Levy2014}; this violation was later shown to be beyond the order of the 
employed approximation, thus only apparent \cite{Trushechkin2016}. Related discussions date back to 2002
\cite{Capek2002,Novotny2002}. Furthermore, a recent paper has shown that the local master equation reconciles with the laws of thermodynamics when analysing a suitable associated collisional model \cite{DeChiara2018a}. Connections between the local evolution of an open system and its thermodynamic microscopic model were previously addressed in Refs.~\cite{Barra2015,PhysRevX.7.021003}. The investigation of the local vs global problem 
in different scenarios is nowadays quite active 
\cite{Santos2014,PhysRevA.90.063815,Manrique2015,Purkayastha2016,Santos2016,Decordi2017,
Gonzalez2017,Hofer2017,Stockburger2017,Hewgill2018,DeChiara2018a,Mitchison2018,PhysRevA.98.052123,seah,raja2018thermodynamic}. 
For instance the failure of the local approach when studying 
two coupled qubits is claimed in Refs. 
\cite{Santos2014,Manrique2015,Decordi2017}. On the contrary, two distinct works 
have tested the validity of the local description applied to the calculation of 
thermodynamics quantities in 
quantum heat engines \cite{Gonzalez2017,Hofer2017}, showing its goodness in a 
quite large range of parameters of the coupling constant, and claiming that the 
global approach fails when the two subsystems are weakly coupled. More precisely,
this assertion is due to a restrictive consideration of the 
global master equation as limited by a \textit{full} secular approximation, 
which also the authors recognise as responsible for the breakdown of the master 
equation. In Ref.~\cite{Gonzalez2017} the possibility for a partial secular 
approximation is also suggested in order to cure such deficiency, and many other 
papers have pointed out why the full secular approximation may not be valid in 
some parameters ranges of different scenarios 
\cite{Wichterich2007,Rivas2010a,Levy2014,Purkayastha2016,PhysRevA.98.052123,seah}.
In the following, we therefore analyze in detail both local and 
global approach and show that the partial secular approximation allows to derive a global master equation that never leads to unphysical results, given that in the limit $\lambda\rightarrow 0$ it coincides with the local master equation. The discussion in Secs.~\ref{sec:nomenclature} 
and~\ref{sec:accuracy} are generally valid also beyond the 2-qubit system, while 
Sec.~\ref{sec:twoQubitLocalGlobal} addresses the validity of the local approach 
and full secular approximation in the specific case of two coupled spins.

\subsection{Setting the nomenclature}
\label{sec:nomenclature}
Let us start by setting a common nomenclature for local and global approach. We 
discuss the case of two subsystems, but generalizations to multipartite systems 
are straightforward. We can thus consider $H_S=H_1+H_2+H_{12}$ with no need of specifying the 
nature of the subsystems and their interaction
 (for two spins we are considering $H_j=\omega_j/2\,\sigma_j^z$, and 
$H_{12}=\lambda\sigma_1^x\sigma_2^x$). We now recall the local and global master 
equations for an open quantum systems.

\paragraph{Local master equation} The local approach is an approximation that 
consists in calculating the jump operators in equation~\ref{eqn:jumpOp} using as free 
system Hamiltonian $H_S^{local}=H_1+H_2$, i.e. neglecting the interaction 
between the subsystems when computing the effects of the environment. This 
clearly leads to two separate sets of local jump operators which (non-trivially) 
act only on the first or second subsystem. If a full secular approximation is 
applied, the direct coupling between the subsystems only appears in the 
commutator $[H_S,\rho_S]$ of the Bloch-Redfield master  
equation~\ref{eqn:BlochRedfieldSchrodinger}, thus it only influences the unitary part 
of the evolution. Intuitively, the local approach is expected 
to provide us with a valid approximated master equation only when the coupling 
constant between the subsystems is sufficiently small \cite{Wichterich2007,PhysRevLett.114.220601,Trushechkin2016,Gonzalez2017,Hofer2017,DeChiara2018a}. 

\paragraph{Global master equation} The global approach consists in considering 
the full (exact) system Hamiltonian, interacting term included, when calculating the 
jump operators. Hence, the global master equation is the Bloch-Redfield one 
 \ref{eqn:preSecular} without further approximations. The jump operators 
appearing on the right term of the equation are not local anymore, i.e. since 
they are obtained after the diagonalization of $H_S$, they can act on both the 
first and the second subsystem. The global master equation is in general
more precise than the local one, as the latter relies on a further 
approximation. It might however be too involved to be solved, due to the 
non-locality of the jump operators \cite{Gonzalez2017}. To simplify its form, one may rely on the 
standard secular approximation, provided that the condition in 
equation~\ref{eqn:secularAppCond} is fulfilled.
 
In the recent literature it is sometimes used the term ``global master equation'' to 
indicate the result of the global approach described 
above \textit{and} after having performed an indiscriminate full secular 
approximation. We believe that 
this nomenclature may lead to 
confusion: per se, the fact of being ``global'', i.e. to lead to jump operators 
which act jointly on both the subsystems, is not related to the secular 
approximation. 
The reported appearance of unphysical currents is not due to the non-locality of the 
jump operators (global approach), being instead the result of the 
indiscriminate application of the full secular approximation when only  the 
partial one was justified \cite{Wichterich2007,Gonzalez2017,Hofer2017}. The 
attempt to apply the full secular approximation for any different frequencies 
$\omega'\neq\omega$, even when the condition in equation~\ref{eqn:secularAppCond} is not fulfilled, leads to inconsistencies: all the 
approximations listed in Sec.~\ref{sec:BlochRedfield} are indeed valid in 
well-defined parameter regimes.
From the formal derivation in the previous section, 
in the framework of Born-Markov approximations,
a global master equation with a justified \textit{partial} secular approximation 
is in general more accurate (or less approximated) than any local one. 
In reference \cite{Gonzalez2017} this was also 
suggested and named partial Markovian Redfield master equation. An important exception is a master equation derived in the
singular-coupling limit  \cite{BreuerPetruccione,Hofer2017}, that would lead to a local master equation. In
this particular case, the global  master  equation  with partial 
secular  approximation, even if accurate, would be unnecessarily more
complicate than the local one. For instance, this is the case when
addressing a Markovian scenario with very high temperature in which the
autocorrelation functions of the bath decay faster than the system
itself \cite{Mitchison2018}. 

One  may  argue  that  the  full  secular  approximation  is anyway 
preferable to the partial one, since it is generally 
introduced to obtain a GKLS master equation \cite{Hofer2017} such 
as the one in equation~\ref{eqn:masterEqFullSecular}, free from any unphysical 
behavior. It is indeed known that the Bloch-Redfield  
equation~\ref{eqn:preSecular} may in some cases violate the positivity of the 
dynamical map \cite{Benatti2005}. However, if the full secular approximation 
is not well justified from a microscopic model (because equation~\ref{eqn:secularAppCond} does not apply)
then a global full secular master equation needs to be considered as a phenomenological one,
as the correspondence with the microscopic model is lost.
On the other hand, as the approximations we have performed to obtain 
equation~\ref{eqn:preSecular} are correct up to the order of the remainders, 
unphysical departures are expected to be consistently small \cite{Hartmann2019}. 
For an in-depth discussion about Bloch-Redfield equation and complete positivity we refer the reader 
to the broad literature 
on the topic 
\cite{Dumcke1979,Suarez1992,Gaspard1999,Wilkie2001,Benatti2003,Benatti2005,
Ishizaki2009,Argentieri2014,Jeske2015,
Purkayastha2016}.

For our purpose, we conclude stressing that 
the GKLS form of the master equation is also guaranteed by the correct application of the partial 
secular approximation, as recently discussed in Refs.~\cite{Cresser2017a,Farina2019}.
This is therefore a preferable approach, being well related to a microscopic model instead of being phenomenological.

\subsection{Accuracy of the local master equation}
\label{sec:accuracy}
In order to assess the accuracy of the local master equation, let us write the 
interaction Hamiltonian as $H_{12}=\lambda V$, where $\lambda$ is a ``small'' 
parameter which we consider as a perturbation order, i.e. $\lambda\ll 1$. The 
system Hamiltonian reads $H_S=H_{1}+H_2+H_{12}=H_{S}^{local}+\lambda V$. 
Following Ref.~\cite{Trushechkin2016}, we apply standard perturbation theory to 
find the zero-th order eigenvectors and eigenvalues\footnote{For simplicity, we 
assume that there are no degenerate eigenvalues. If this is not the case, one 
has to diagonalize the interaction Hamiltonian in the degenerate subspace, 
according to degenerate perturbation theory \cite{Sakurai}, and still recovers 
the results we are going to present in the following. In 
Sec.~\ref{sec:twoQubitLocalGlobal} we will present a case in which a degeneracy 
may occur, and discuss when the local approach is still valid.}. Within the 
assumption of not-degenerate system Hamiltonian, we write the eigenvalues as the 
infinite perturbation 
expansion~\cite{Sakurai}:
\begin{equation}
\label{eqn:perturbationTheory}
\begin{aligned}
&E_n=E_n^{(0)}+\lambda E_n^{(1)}+\lambda^2 E_n^{(2)}+...\\
&\ket{e_n}=\ket{e_n^{(0)}}+\lambda\ket{e_n^{(1)}}+\lambda^2\ket{e_n^{(2)}}+...
\end{aligned}
\end{equation}
where $E_n^{(0)}$ and $\ket{e_n^{(0)}}$ are respectively the eigenvalues and 
eigenvectors of the unperturbed Hamiltonian, in our case of the local 
Hamiltonian $H_S^{local}$. A jump operator equation~\ref{eqn:jumpOp} using the 
expansions in equation~\ref{eqn:perturbationTheory} will read
\begin{eqnarray}
\label{eqn:perturbedJump}
A_\beta(\omega)&=\sum_{E'_n-E_n=\omega}\ket{e_n}\bra{e_n}{A_\beta}\ket{e'_n}
\bra{e'_n}\nonumber\\
&=\underbrace{\sum_{E'_n-E_n=\omega}\ket{e^{(0)}_n}\bra{e^{(0)}_n}{A_\beta}
\ket{e'^{(0)}_n}\bra{e'^{(0)}_n}}_{A^{(0)}_\beta(\omega^{(0)})}+O(\lambda),
\end{eqnarray}
where $A^{(0)}_\beta(\omega^{(0)})$ are the local jump operators appearing in 
the local master equation.

While Ref.~\cite{Trushechkin2016} considers the GKLS master equation in full 
secular approximation, here we derive the Bloch-Redfield local master equation. 
Inserting equation~\ref{eqn:perturbedJump} in equation~\ref{eqn:preSecular} and coming 
back to the Schr\"odinger picture we obtain:
\begin{eqnarray}
\label{eqn:BlochRedfieldPerturbed}
\frac{d}{dt}\rho_S(t)=&-i[H_S^{local}+\lambda V,\rho_S(t)]\nonumber\\
&+\sum_{\omega^{(0)},\omega'^{(0)}}\sum_{\beta,\beta'}\Gamma_{\beta\beta'}
(\omega^{(0)})\Big(A_{\beta'}^{(0)}(\omega^{(0)})\rho_S(t)(A^{(0)}
_\beta)^\dagger(\omega'^{(0)})\\
&\qquad\qquad\qquad\qquad\qquad-(A^{(0)}_\beta)^\dagger(\omega'^{(0)})A^{(0)}_{
\beta'}(\omega^{(0)})\rho_S(t)\Big)\nonumber\\
&+h.c.+O(\mu^2\lambda)+o(\mu^2),\nonumber
\end{eqnarray}
where $\omega^{(0)}$ is the frequency given by differences of the unperturbed 
eigenvalues $E^{(0)}_n-E^{(0)}_m$. The error we are making by 
employing the local master equation is of the order of $O(\mu^2\lambda)$, since 
the leading order of the master equation after the Born approximation is 
$O(\mu^2)$. Moreover, we explicitly write the order of the remainder after the 
Born-Markov approximations $o(\mu^2)$, to stress that there are already some 
neglected terms which may be larger than the error given by the local master 
equation $O(\mu^2\lambda)$. 

If we employ the local master equation to compute physical quantities, clearly 
we can resolve them only up to the order of $O(\mu^2)$. Any quantity of the 
order of $O(\mu^2\lambda)$ or smaller, then is null in the framework of the 
local approach. This is the reason why the violation of the second law of 
thermodynamics \cite{Levy2014} is only an apparent one \cite{Trushechkin2016}, 
given that it is of the order of $O(\mu^2\lambda^2)$.

\subsection{Local vs global approach for two coupled qubits}
\label{sec:twoQubitLocalGlobal}
We will now address the local vs global comparison focusing on the case of two 
spins as in equation~\ref{eqn:freeSystemHamInitRen}, with ``local'' Hamiltonian 
$H_S^{local}=\frac{\omega_1}{2}\sigma_1^z+\frac{\omega_2}{2}\sigma_2^z$ and 
interaction $H_{12}=\lambda \sigma_1^x\sigma_2^x$. As discussed in the previous 
sections, the global master equation with partial secular approximation is 
always valid up to the errors induced by the Born-Markov approximations. On the 
contrary, the local master equation and the full secular approximation are 
accurate only for some parameter regimes, which we will investigate starting 
from the derivation of the master equation. We recall the fact that the local approach is always valid in the singular-coupling limit, which is more restrictive.

We will first present the local master equation for two coupled qubits, studying the ranges of parameters in which each approximation works, 
in the presence of common or separate baths, and then show how the 
difference between master equations emerges through some illustrative examples.

\subsubsection{Local master equation}
The local approach is valid in the case in which $\lambda\ll 1$. In order to derive the master equation we must find the jump operators, thus the eigenvalues and eigenvectors of the Hamiltonian.
Let us start from the situation where the local Hamiltonian is not degenerate, i.e. $\lambda\ll\omega_-$. In this case we have the following eigenvalues:
\begin{equation}
\label{eqn:eigenValPert}
\begin{aligned}
&E_0=-\frac{\omega_+}{2}+O(\lambda),\qquad & E_1=-\frac{\omega_-}{2}+O(\lambda),\\
&E_2=+\frac{\omega_-}{2}+O(\lambda),\qquad & E_3=+\frac{\omega_+}{2}+O(\lambda),\\
\end{aligned}
\end{equation}
with corresponding eigenvectors
\begin{equation}
\label{eqn:eigenVecPert}
\begin{aligned}
&\ket{e_0}=\ket{00}+O(\lambda),\qquad&\ket{e_1}=\ket{01}+O(\lambda),\\
&\ket{e_2}=\ket{01}+O(\lambda),\quad\,&\ket{e_3}=\ket{11}+O(\lambda).\\
\end{aligned}
\end{equation}
Equations~\ref{eqn:eigenValPert} and~\ref{eqn:eigenVecPert} are the zero-th order expressions for the infinite perturbative expansion in equation~\ref{eqn:perturbationTheory}. By inserting them in equation~\ref{eqn:BlochRedfieldPerturbed}, we see that the local master equation is exactly the master equation~\ref{eqn:masterEqNoDir} found for decoupled qubits, but with free system Hamiltonian $H_S$ including a coupling term, i.e. $H_S$ given by equation~\ref{eqn:systemHamDir} instead of equation~\ref{eqn:systemHamNoDir}. So, the difference between the local and the global master equation is of the order of $O(\mu^2\lambda)$.

In the degenerate regime  $\omega_-=0$, there is an apparent freedom in the choice of the basis with respect to which the perturbative expansion must be performed, as any linear combination of $|e_1\rangle$ and  $|e_2\rangle$ could in principle be selected. This apparent freedom is actually removed by  the interaction Hamiltonian. For instance, in the case of separate baths, deriving the master equation starting from any possible choice of the basis would in any case lead to local jump operators, as in the absence of degeneracy.  Working near degeneracy, that is, assuming $\lambda\gg\omega_-\neq 0$, we would have for instance $\ket{e_2}=1/\sqrt{2}(\ket{01}+\ket{10})+O(\lambda)$, but the aforementioned selection rule would still  apply and the final master equation would not change. 

\subsubsection{Comparison between master equations}
\label{sec:comparison}
In this section we will investigate the limits of validity of the full secular 
approximation and of the local master equation depending on the parameters of 
the system and show some 
examples of the different system dynamics generated by them. In Sec.~\ref{sec:physicalQ} 
we will provide further examples focused on physical quantities computed through distinct 
master equations. The results are summarized in Table~\ref{tab:tabella} 
presented in the concluding remarks.
The parameters we vary in order to study each master equation are the 
qubit-qubit coupling constant $\lambda$ and the detuning between the qubits 
$\omega_-$, as can be seen in Table~\ref{tab:tabella}. All the remaining 
parameters will be  fixed as follows, paying attention to the conditions for the 
Born-Markov approximations: 
\begin{itemize}
\item We choose as qubit-bath coupling constant $\mu=10^{-2}$. This means that 
the timescale of the evolution of the system will be $\tau_R=O(\mu^{-2})=10^{4}$. 
This quantity is important to check the validity of each approximation, as shown 
in Table~\ref{tab:tabella}. The remainder given by the Born approximation will 
be, according to equation~\ref{eqn:integroDiffEqBis}, of the order of 
$O(\mu^3)=10^{-6}$.
\item Both the common and the separate baths will have an Ohmic spectral 
density, i.e. 
\[J(\omega)=\omega\frac{\Omega^2}{\Omega^2+\omega^2},
\] where $J(\omega)$ is defined in equation~\ref{eqn:spectralDensity} and $\Omega$ is 
a cutoff frequency which we have set as large as $\Omega=20$. Regarding the 
inverse of the temperature of each bath, we have chosen $\beta^{(c)}=1$, 
$\beta^{(l_1)}=1$, $\beta^{(l_2)}=0.1$. An unbalance between the local baths is 
important in quantum thermodynamics, in order to study the heat transport 
between them. We finally have to check that these baths with the chosen 
temperatures satisfy the condition for the Markov approximation 
equation~\ref{eqn:markovCondition}. The timescale of the decay of the bath 
autocorrelation functions for an Ohmic spectral density reads 
\cite{BreuerPetruccione} $\tau_B=\textnormal{Max}\{\Omega^{-1},\beta/2\pi\}$, 
and $\tau_R=O(\mu^4)$, therefore $\tau_B\ll\tau_R$ and the Markov approximation 
is valid for our choice of parameters.
\item We choose as initial state of the system the factorized state 
$\rho_0=\rho_{OV}\otimes\rho_{OV}$, where $\rho_{OV}$ is the completely 
overlapped state 
$\rho_{OV}=1/2(\ket{0}\bra{0}+\ket{0}\bra{1}+\ket{1}\bra{0}+\ket{1}\bra{1})$.
\end{itemize}

We will evaluate the dynamics using four different master equations, namely the 
global master equation in partial secular approximation (GP), which we will consider as the most correct one according to the discussion in Sec.~\ref{sec:BlochRedfield}, the global master equation in full secular approximation (GF), the local 
master equation in partial secular approximation (LP), and the local master equation 
in full secular approximation (LF). We remind that here we are using ``partial secular approximation'' to refer to a master equation in which we keep the cross terms which, in some scenario, may be slow-rotating and not negligible, even in regimes in which such terms may actually be eliminated. Each master equation leads to a (a priori) 
different evolution. For simplicity, we focus on three different figures of 
merit. The first one is the mean value of $\sigma_1^z$ as a function on time, 
i.e. $\langle\sigma_1^z(t)\rangle$. The second is the fidelity 
\cite{NielsenChuang} between the state obtained through the global master 
equation with partial secular (most accurate one) and a state computed with another 
master equation, i.e. $\mathcal{F}(\rho_{GP}(t),\cdot)$, as a function on time. 
The third and last figure of merit is the steady state of the system, i.e. the 
state obtained for $t\rightarrow\infty$. Notice that, while the fidelity is 
quite a general and reliable 
indicator, both the population of the first qubit and the steady state may not 
display differences between two master equations, even if the latter are 
substantially different. 

For convenience, we first study the scenario with separate baths only, and then 
the one in the presence of a common bath, addressing in both cases the local and global master equation separately, and focusing 
on dissipative couplings with the environments, as we do not expect any qualitative difference if we also add dephasing baths.

\subsubsection*{Separate baths}
\begin{figure}
\centering
\includegraphics[scale=0.28]{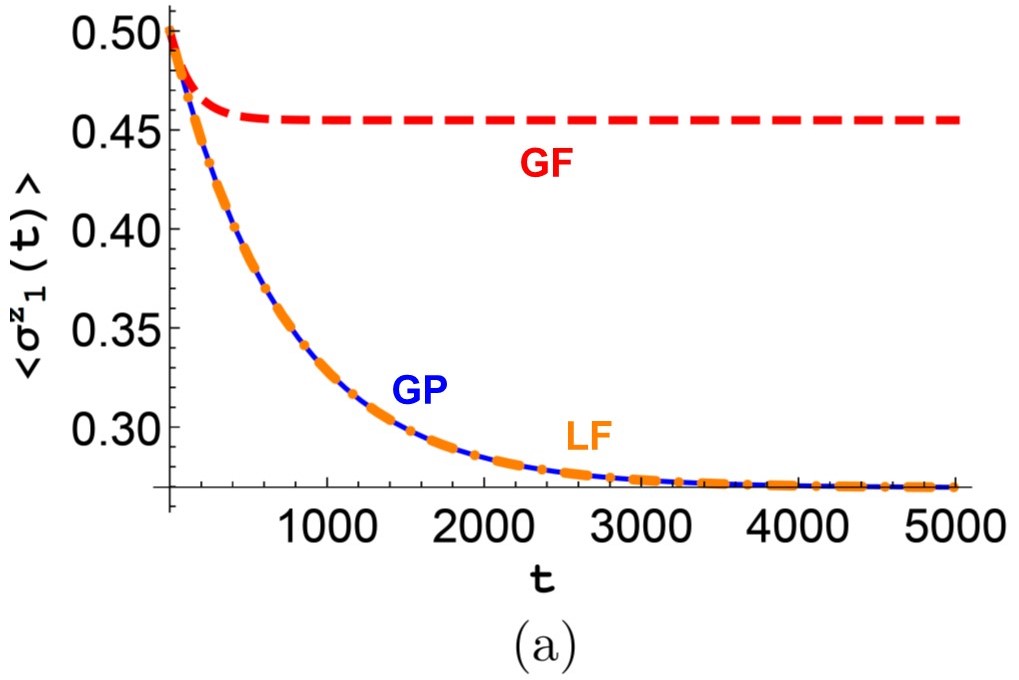}
\includegraphics[scale=0.28]{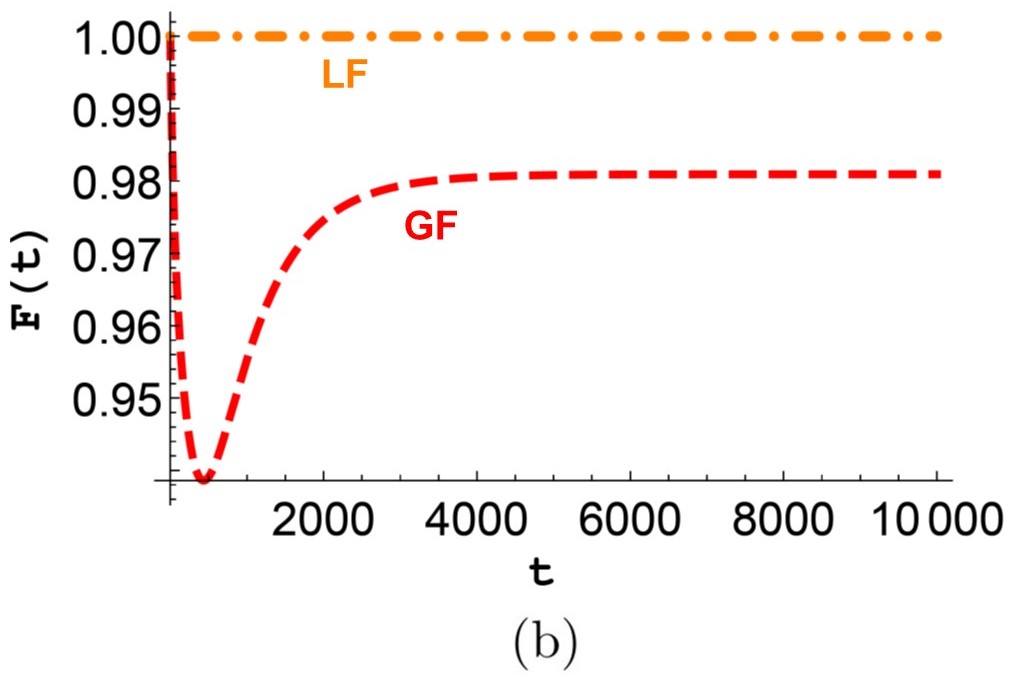}
\includegraphics[scale=0.28]{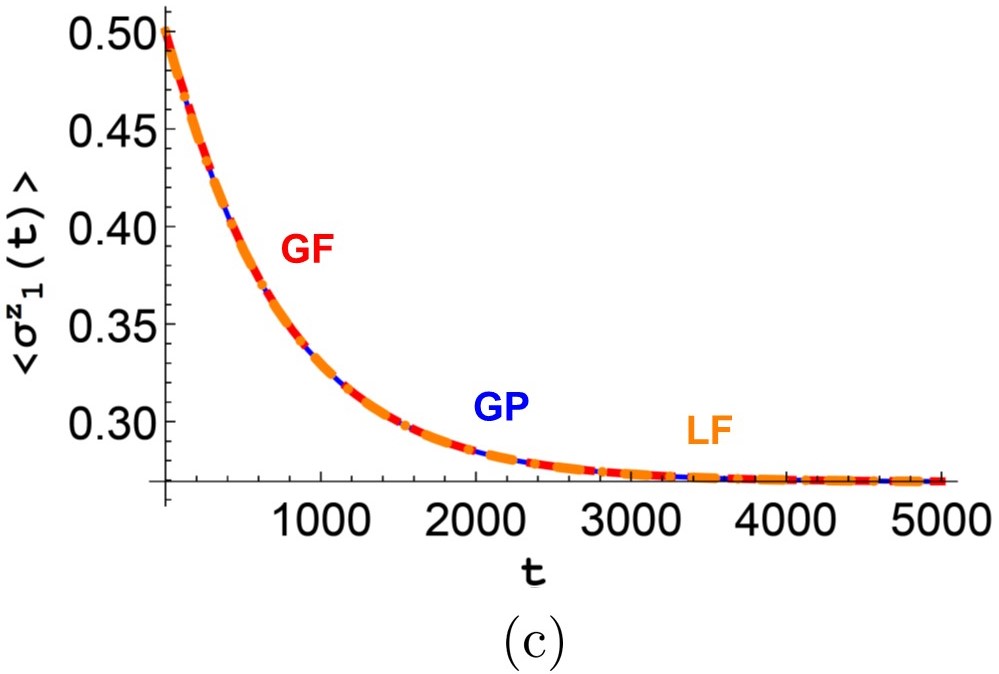}
\includegraphics[scale=0.28]{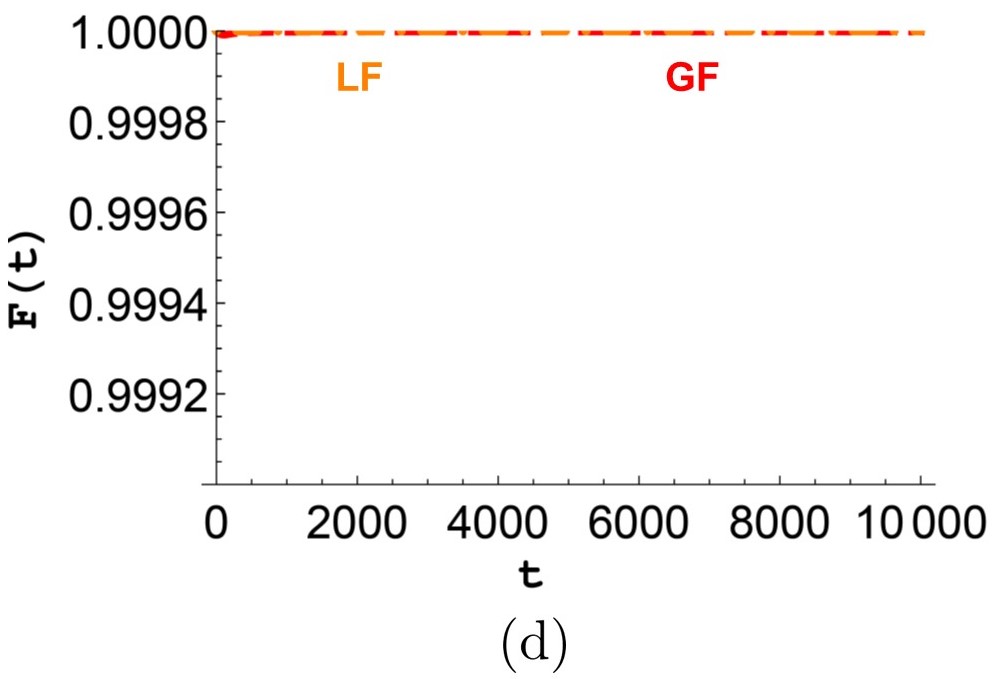}
\caption{Comparison between master equations (m.e.) in the 
presence of separate baths only, for 
coupled spins ($\lambda=10^{-4}$) with identical frequencies.  
(a) and (b): $\omega_-=0$. (c) and (d): $\omega_-=0.01$. All the other parameters have been 
set according to the discussion in Sec.~\ref{sec:twoQubitLocalGlobal}. 
(a) and (c): population of the first qubit as a function of time, 
according to the global m.e. with partial secular approximation GP (solid blue), 
the global m.e. with full secular approximation GF (dashed red), and the local m.e. with full secular approximation  LF (dot-dashed orange). (b) and (d): 
fidelity between the state obtained through the global m.e. with partial secular 
approximation and respectively global m.e. with full secular approximation 
$\mathcal{F}(\rho_{GP}(t),\rho_{GF}(t))$ (dashed red) and local m.e. 
with full secular approximation $\mathcal{F}(\rho_{GP}(t),\rho_{LF}(t))$ (dot-dashed 
orange). Note that local full (LF) coincides with local partial (LP) as explained in the main text.}
\label{fig:NoDet}
\end{figure}

\begin{figure}
\centering
\includegraphics[scale=0.3]{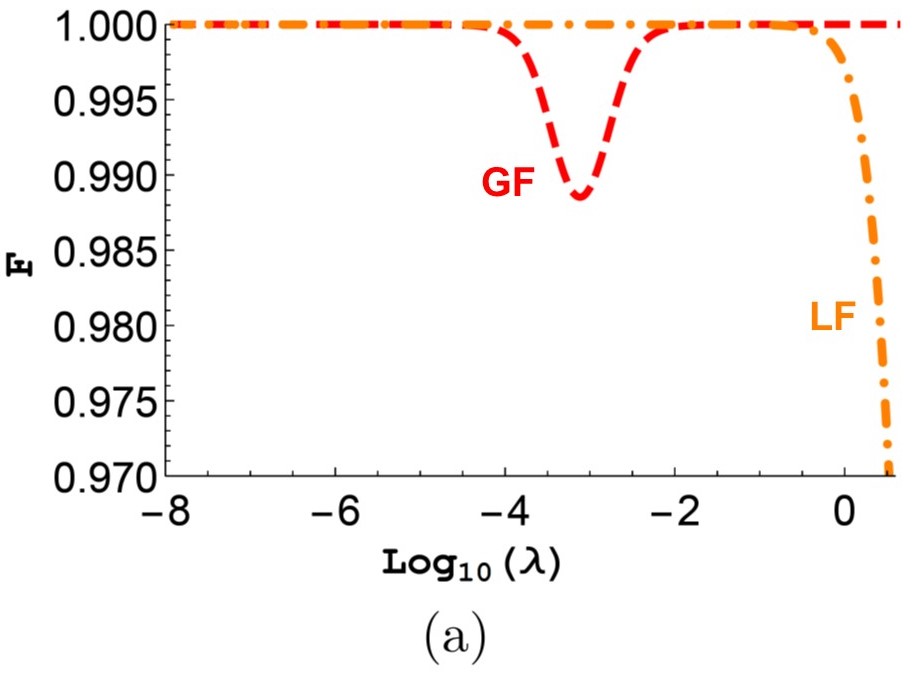}
\includegraphics[scale=0.3]{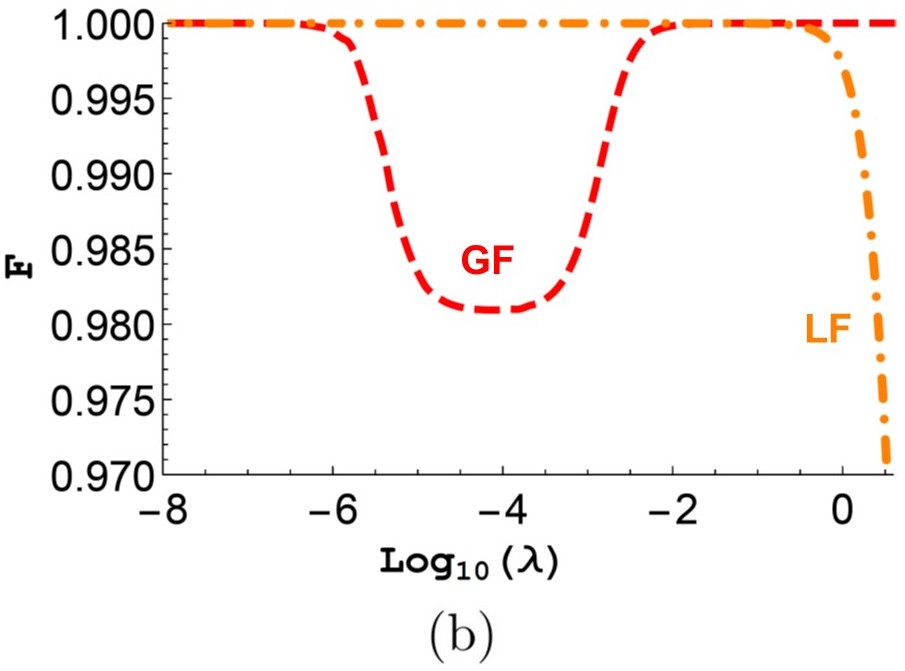}
\caption{ Fidelity between the steady state obtained through the 
global master equation (m.e.) with partial secular approximation and 
respectively global m.e. with full secular approximation $\mathcal{F}(\rho_{GP},\rho_{GF})$ (dashed red) and local 
m.e. with full secular approximation 
$\mathcal{F}(\rho_{GP},\rho_{LF})$ (dot-dashed orange), as a function of 
the qubit-qubit coupling constant $\lambda$. The qubits are interacting with two 
separate thermal baths at inverse temperature $\beta^{(l_1)}=1$, 
$\beta^{(l_2)}=0.1$, and the qubit-bath coupling constant reads $\mu=10^{-2}$. 
(a): $\omega_-=10^{-3}$. (b): $\omega_-=10^{-5}$. We anticipate the regimes of validity summarized in 
Table~\ref{tab:tabella}. Note that local full (LF) coincides with local partial (LP) as explained in the main text.
}
\label{fig:steady}
\end{figure}

\begin{itemize}
\item In the case of separate baths, the local master equation presents local 
dissipators on each qubit, and the interaction between the subsystems only comes 
into play in the unitary part of the evolution. This means that the full secular 
approximation always coincides with the partial one and is  valid for any value of the detuning $\omega_-$, provided that  $\lambda\ll 1$, which fixes the validity of the local master equation.

\item The full secular approximation with global approach may break down for some range of 
values. Indeed, the global approach makes use of the basis of eigenmodes to 
build the jump operators, which is composed of entangled states, therefore a 
single local bath coupled to $\sigma_1^x$ induces dissipation on the second 
qubit as well, and the secular approximation comes into play. Let us look at the 
jump frequencies presented in equation~\ref{eqn:jumpFrequencies}. We have to identify 
the frequency differences which may be comparable with $\tau_R^{-1}=O(\mu^2)$, always avoiding the singular parameter choice expressed in 
equation~\ref{eqn:conditionCrossing}. Critical cases are:
\begin{equation}
\label{eqn:energyDiff}
\omega_{II}-\omega_I=\omega_{IV}-(-\omega_{IV})=2(\omega_{IV}-\omega_0)=2\sqrt{
\lambda^2+\omega_-^2/4}.
\end{equation}
If the qubits have small detuning, i.e. $\abs{\omega_-}\ll 1$, and the 
qubit-qubit coupling constant is small as well, $\lambda\ll 1$, the secular approximation on these frequencies 
equation~\ref{eqn:secularAppCond} does not apply 
 if for instance 
$\abs{\omega_{II}-\omega_I}\lessapprox\mu^{2}$.
Still, if the basis of eigenvectors of 
the system Hamiltonian is quasi-local, in the sense that it is well described by 
the states in equation~\ref{eqn:eigenVecPert}, the cross terms between the qubits 
arising with separate baths are very small, of the order of $O(\mu^2\lambda)$. 
Thus we can neglect them, and the full secular approximation is still valid even 
in the global case. This final condition of validity reads $\lambda\ll\omega_-$, 
highlighting the importance of the ratio between detuning and qubit-qubit 
coupling constant. 
\end{itemize}

We show an example in figure~\ref{fig:NoDet}, considering $\langle\sigma_1^z(t)\rangle$ and the fidelity 
between evolved states as a function of time, and in figure~\ref{fig:steady}, 
focusing on the steady state.
Being the baths separate, the local master equation with full secular 
approximation coincides 
with the partial secular one. In figure~\ref{fig:NoDet}, since $\lambda$ is very small, the local approach 
provides a reliable description of the dynamics independently of the value of 
the detuning. On the contrary, for identical qubits (figure~\ref{fig:NoDet}a and figure~\ref{fig:NoDet}b, with $\omega_-=0$), the global 
master equation with full secular approximation fails, while this approximation in the global approach
is justified  for $\lambda\ll\omega_-$ (figure~\ref{fig:NoDet}c and figure~\ref{fig:NoDet}d), despite the 
detuning being small. Looking at the stationary states and also allowing for baths at different temperatures,
we show in figure~\ref{fig:steady} the predicted parameters regimes of failure of the local master 
equation and of the full secular approximation in the global one.
While $\lambda\ll\omega_-$, both the approaches are 
reliable, but as soon as $\lambda$ gets close to $\omega_-$ the GF fails; 
this clearly starts from smaller values of $\lambda$ on the figure~\ref{fig:steady}
left than on the right, since in the former case the detuning is smaller. As far 
as $\lambda$ increases 
toward $1$, the global approach with full secular recovers validity, since it 
fulfills the condition for the full secular approximation. As $\lambda$ becomes 
of the order of the qubit frequency $O(1)$, the local m.e. loses reliability.

\subsubsection*{Common bath} 
\begin{figure}[t]
\centering
\includegraphics[scale=0.28]{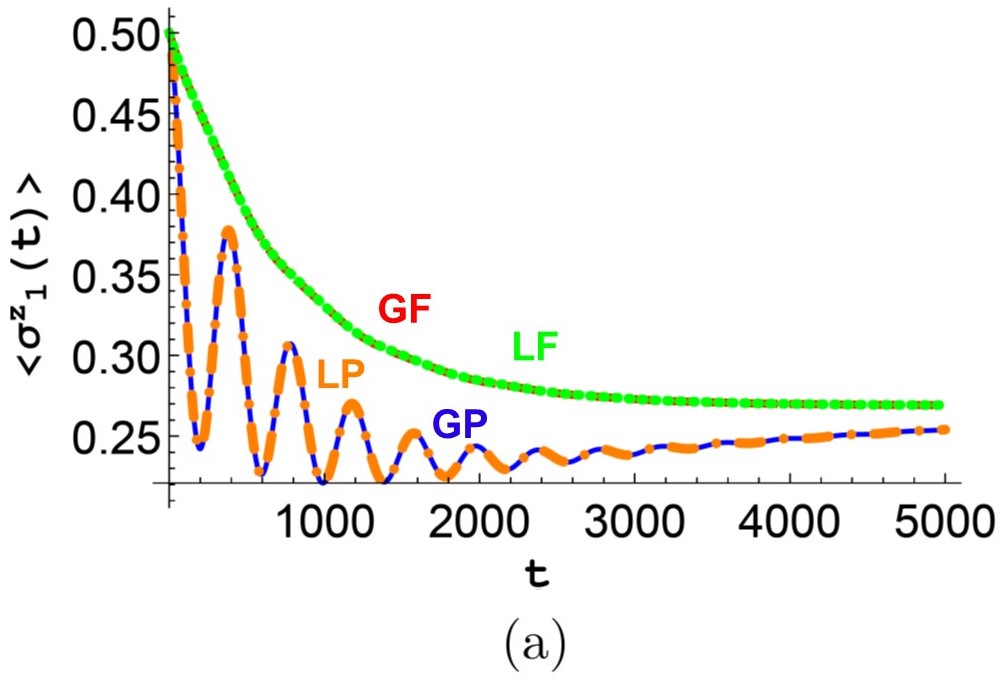}
\includegraphics[scale=0.28]{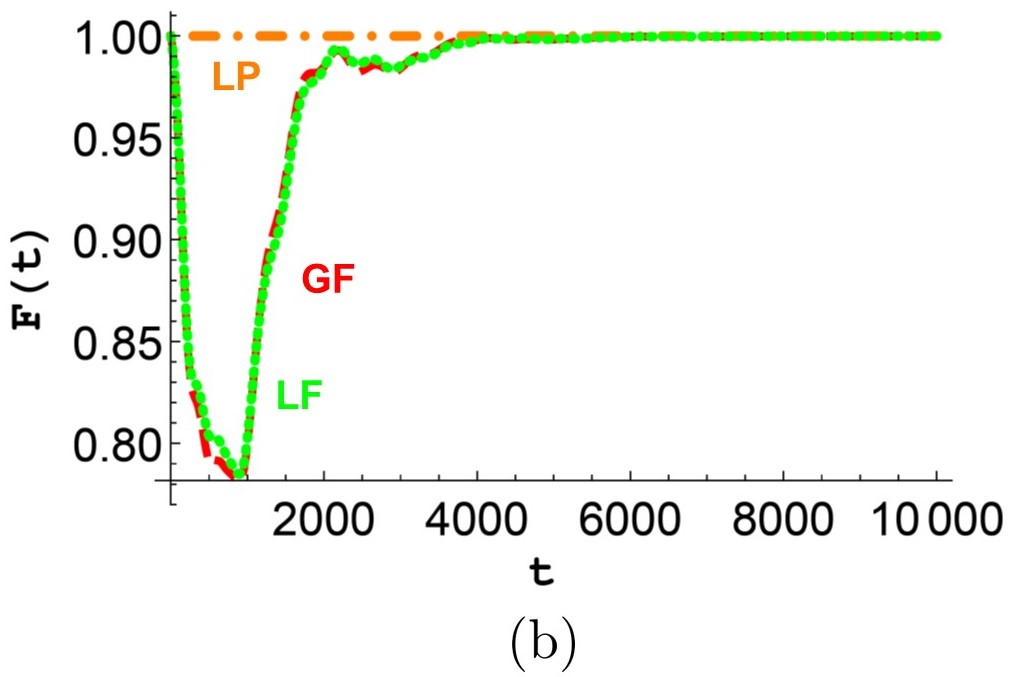}
\includegraphics[scale=0.28]{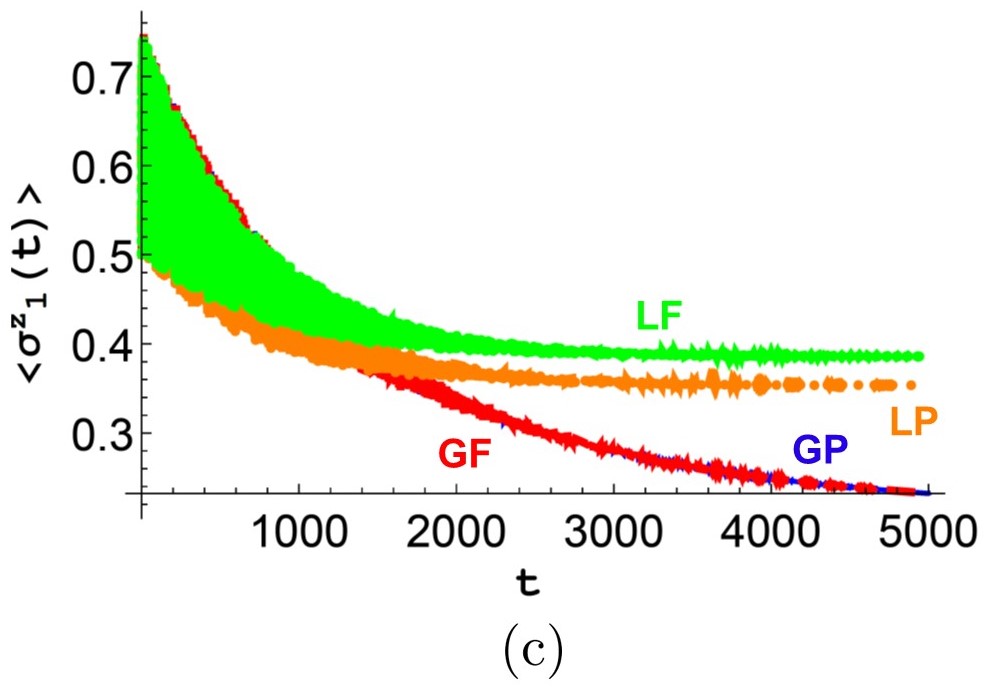}
\includegraphics[scale=0.28]{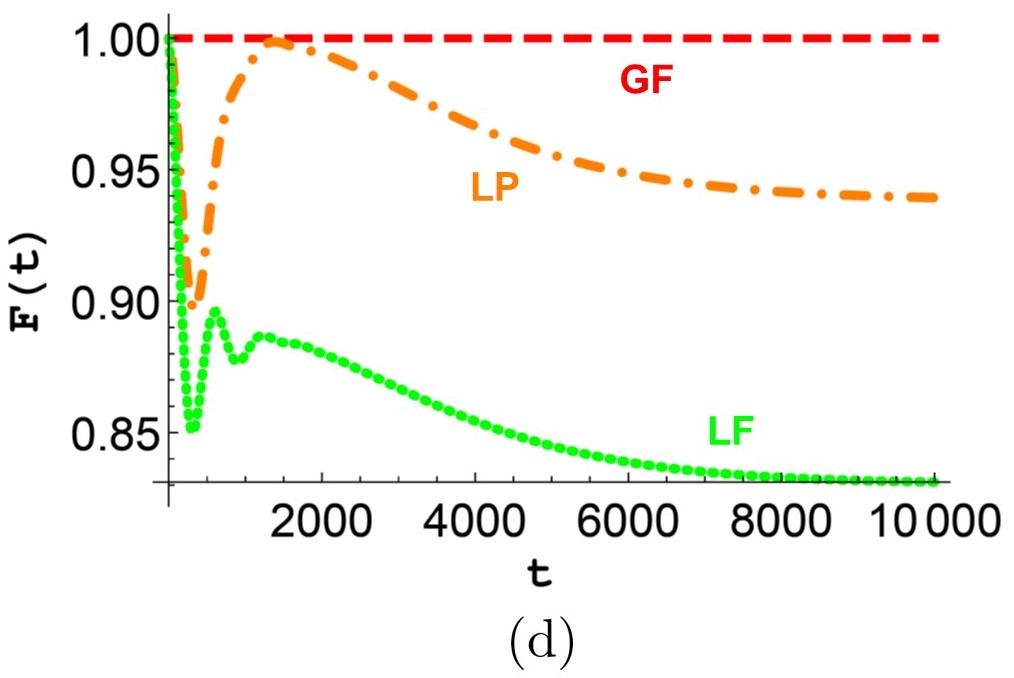}
\caption{ Comparison between master equations (m.e.) in the 
presence  of a common bath, for weak and strong coupling between qubits. (a) and (b): $\omega_-=0.01$, $\lambda=10^{-4}$. (c) and (d): 
$\omega_-=0.01$, $\lambda=1$. All the other parameters have been set according 
to the discussion in Sec.~\ref{sec:twoQubitLocalGlobal}. (a) and (c): 
population of the first qubit as a function of time, according to the global 
m.e. with partial secular approximation GP (solid blue), the global m.e. with full 
secular approximation GF (dashed red), the local m.e. with partial secular 
approximation LP (dot-dashed orange), and the local m.e. with full secular 
approximation LF (dotted green). (b) and (d): fidelity between the state 
obtained through the global m.e. with partial secular approximation and 
respectively global m.e. with full secular approximation (dashed red), local 
m.e. with partial secular approximation (dot-dashed orange) and local m.e. with full secular approximation (dotted green).}
\label{fig:ComPar}
\end{figure}
\begin{itemize}
\item If the bath is common, in general the local master equation does not 
display local dissipators. Indeed, if the detuning $\omega_-$ is small, i.e. not 
way larger than $\tau_R^{-1}=O(\mu^2)$, we obtain cross terms in the master 
equation which have the effect of exchanging excitations between the qubits (see 
equation~\ref{eqn:masterEqNoDir}). Therefore, the full secular approximation for the 
local master equation is valid only if $\omega_-\gg\mu^2$. This means that the 
claim about the goodness of an indiscriminate full secular approximation when 
following the local approach is not general, but limited to the separate baths 
scenario only. Of course, the local master equation is accurate only if 
$\lambda\ll 1$.
\item For the global master equation in the presence of a common bath, the same 
discussion about the case with separate baths hold, with the only difference 
that a local basis ($\lambda\gg\omega_-$) does not allow us to perform the full 
secular approximation anymore. Therefore, the condition for the global master 
equation with full secular approximation reads $\omega_-\gg O(\mu^2)$ or 
$\lambda\gg O(\mu^2)$. 
\end{itemize}

Figure~\ref{fig:ComPar} shows some relevant examples through $\langle\sigma_1^z\rangle$ 
and the fidelity of the states obtained with different master equations compared with the GP: 
if $\lambda$ is very small (first 
row), the local approach with partial secular approximation provides a reliable 
description of the dynamics. On the contrary, both the local and global approach 
with full secular approximation fail, since the detuning is very small as well. 
In the scenario of $\lambda$ being of the order of the qubit frequency (second 
row), the local approach always fails, while the global approach with full 
secular approximation is reliable in spite of the small detuning, since 
$\lambda\gg\mu^2$.

\section{Computing physical quantities through distinct master equations}
\label{sec:physicalQ}

In this section we provide some examples of the effect of considering distinct master 
equations on some relevant physical quantities and discuss their accuracy on physical grounds.
We will therefore corroborate the 
mathematical analysis in Sec.~\ref{sec:localGlobal} by suitable 
physical examples. Throughout the section, when not explicitly stated the values of the 
parameters used in the examples are the ones fixed in Sec.~\ref{sec:comparison}.

\subsection{Common bath: entanglement, quantum beats and synchronization}
We will now show how, in the critical regime of small qubit-qubit coupling constant $\lambda$, small detuning and a common bath, the full secular approximation leads to unphysical results, while the partial secular correctly describes several physical phenomena, both in the local and in the global approach. Since we are considering the limit $\lambda\ll 1$, the same considerations hold for the case of uncoupled qubits discussed in Sec.~\ref{sec:noDirect}. In particular, we consider the case addressed in Figs.~\ref{fig:ComPar}a and~\ref{fig:ComPar}b for the coupled case, i.e. $\omega_-=10^{-2}$, $\lambda=10^{-4}$ and the rest of parameters as discussed in Sec.~\ref{sec:comparison}; to include a discussion about the uncoupled case, we consider the scenario of Fig.~\ref{fig:evNoDir}b with $\omega_-=10^{-2}$. These two situations are almost equivalent due to the very small $\lambda$, as can be seen by comparing Fig.~\ref{fig:evNoDir}b and Fig.~\ref{fig:ComPar}a.

\begin{figure}[t]
\centering
\includegraphics[scale=0.285]{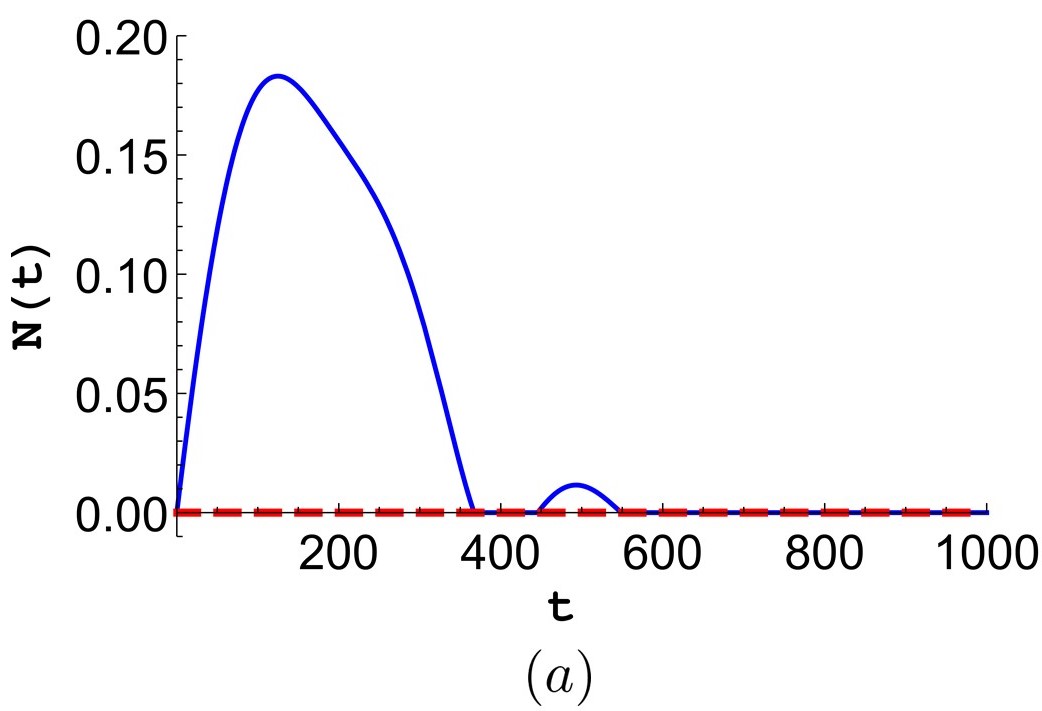}
\includegraphics[scale=0.285]{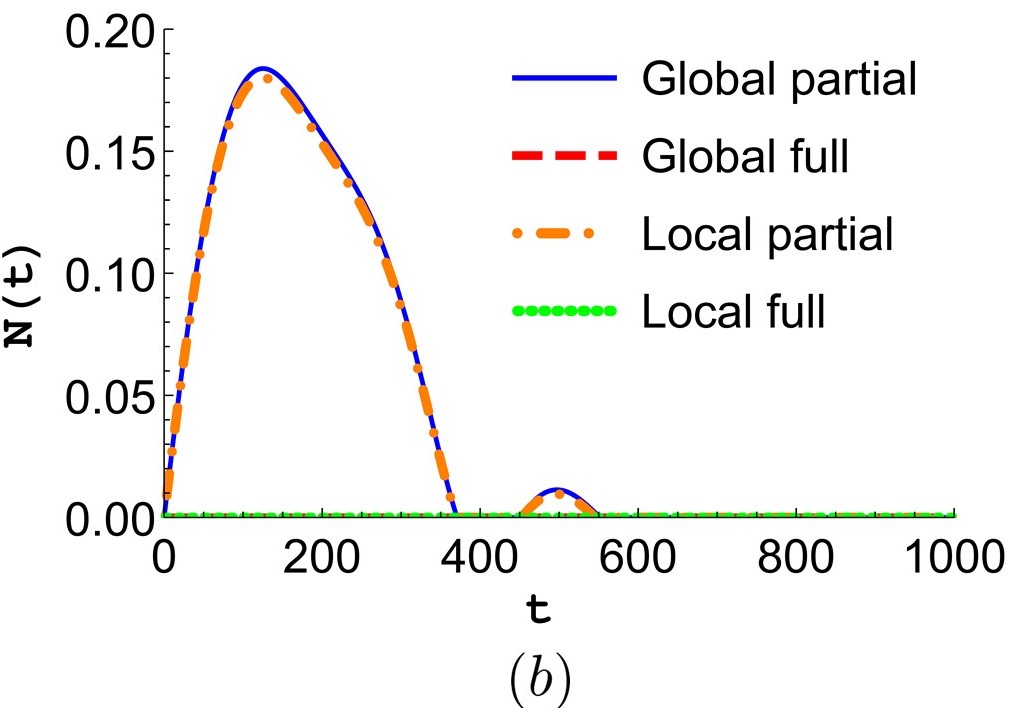}
\caption{Negativity as a function of time in the presence of a common bath with inverse temperature $\beta^{(c)}=1$, for $\omega_-=10^{-2}$ and no direct coupling (a) or $\lambda=10^{-4}$ (b). We confront the value obtained through the global master equation (m.e.) with partial secular approximation (solid blue), the global m.e. with full
secular approximation (dashed red), the local m.e. with partial secular
approximation (dot-dashed orange), and the local m.e. with full secular
approximation (dotted green). For both local and global approach, the full secular approximation incorrectly shows no entanglement during the evolution of the qubits, while both the global and local m.e. with partial secular approximation provide the phenomena of entanglement creation, sudden death and sudden birth. Since $\lambda$ is very small, we observe no remarkable difference between the coupled and uncoupled case.}
\label{fig:nega}
\end{figure}

We first focus on the dynamics of entanglement obtained through different master equations. It is indeed well-known that a common bath may generate entanglement between non-interacting qubits immersed in it \cite{aolita2015open}, even when the reservoir is Markovian \cite{Benatti2003a}. This phenomenon has been predicted using diverse methods of obtaining a master equation, such as a coarse-graining procedure \cite{PhysRevA.81.012105} or employing master equations originally derived for quantum optics \cite{PhysRevA.74.024304}. As entanglement measure we choose the negativity $\mathcal{N}$ \cite{aolita2015open}; in the consider case of two qubits, a non-zero value of the negativity is a necessary and sufficient condition to have entanglement. We plot in Fig.~\ref{fig:nega} the negativity as a function of time for both uncoupled and coupled case, which do not show a visible difference  consistently with the very small qubit-qubit coupling constant. 
Both in the local and global approach, the master 
equation with partial secular approximation correctly displays entanglement creation, sudden death and subsequent sudden birth \cite{tanas2011sudden}, that resemble dynamics already observed in similar scenarios \cite{PhysRevA.74.024304}. On the contrary, the full secular approximation completely misses the detection of entanglement, since it does not include a qubit-qubit coupling mediated by the common bath.

As already discussed in Sec.~\ref{sec:noDirect}, another physical phenomenon that the full secular approximation is not able to reproduce are the quantum beats, i.e. the oscillations in the dynamics of the qubit populations that can be observed in Figs.~\ref{fig:ComPar}a and~\ref{fig:evNoDir}b. The quantum beats are known since the studies on superradiance in the eighties \cite{gross1982superradiance}, and appear during the evolution of two slightly-detuned atoms because of the tiny difference between their frequencies in the phase of the emission power \cite{ficek1987quantum}. The full secular approximation does not make the two qubits ``communicate'', and therefore it doesn't detect the detuning and leads to an incorrect smooth decay of the population of the excited state.

Finally, we mention another important physical effect missed by full secular approximation. Quantum synchronization is a paradigmatic phenomenon investigated in disparate fields in the recent years (for a review see \cite{galve2017quantum}). In particular, spontaneous synchronization of the spinning frequencies of two uncoupled qubits in a common bath has been predicted using a Bloch-Redfield master equation without any further secular approximation \cite{Giorgi2013}. Using the master equation ~\ref{eqn:masterEqNoDir} in partial secular approximation we have observed quantum synchronization starting from a time $t\approx 6000$, while the same master equation in full secular approximation never displays synchronization of the qubit frequencies, and it is therefore not suitable to analyze such a phenomenon.

\subsection{Steady state heat current incoming from separate baths}
We now consider the case of two coupled qubits and separate baths addressed 
in Fig.~\ref{fig:steady}: 
for small values of the detuning and varying the coupling constant $\lambda$,
we compute the steady state heat currents coming from the hot and cold reservoir. 
As in Sec.~\ref{sec:comparison}, we assume that the inverse temperatures of the baths 
are $\beta^{(l_1)}=1$ (colder), $\beta^{(l_2)}=0.1$ (hotter). 
If $\rho_\infty$ is the steady state of the open system, 
the heat current from the hotter reservoir is defined as \cite{Alicki1979}:
\begin{equation}
\label{eqn:heatCurrHot}
J_2=\Tr\left(H_S \mathcal{D}_2[\rho_\infty]\right),
\end{equation}
where $H_S$ is the system Hamiltonian equation~\ref{eqn:systemHamDir} and $\mathcal{D}_2$ is the dissipator generated by the hotter bath coupled to the second qubit. Analogously we can define the heat current incoming from the colder reservoir as $J_1=\Tr\left(H_S \mathcal{D}_1[\rho_\infty]\right)$, and since $\rho_\infty$ is the steady state we have $J_1+J_2=0$.
Note that equation~\ref{eqn:heatCurrHot} is widely used in the literature \cite{Alicki1979,Gonzalez2017,Hofer2017}, but cannot be associated to a heat current observable. A definition based on a current observable can be found in Ref.~\cite{PhysRevA.85.032110}, which also discusses some issues regarding its consistency with different master equations.

\begin{figure}[t]
\centering
\includegraphics[scale=0.265]{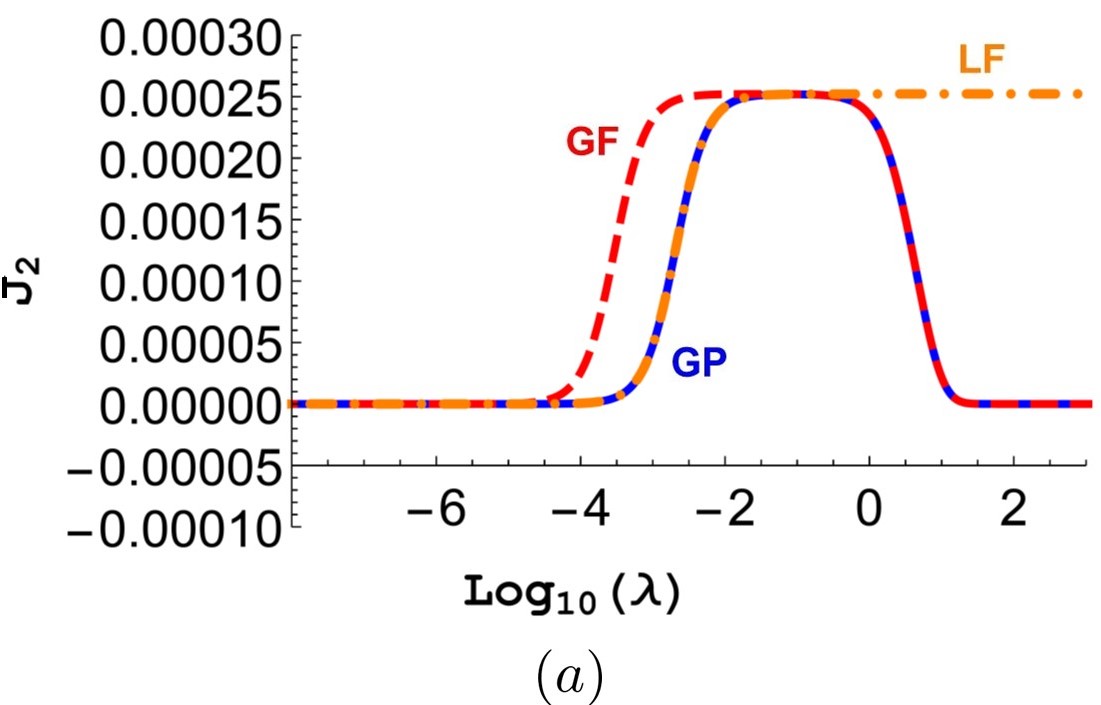}
\includegraphics[scale=0.265]{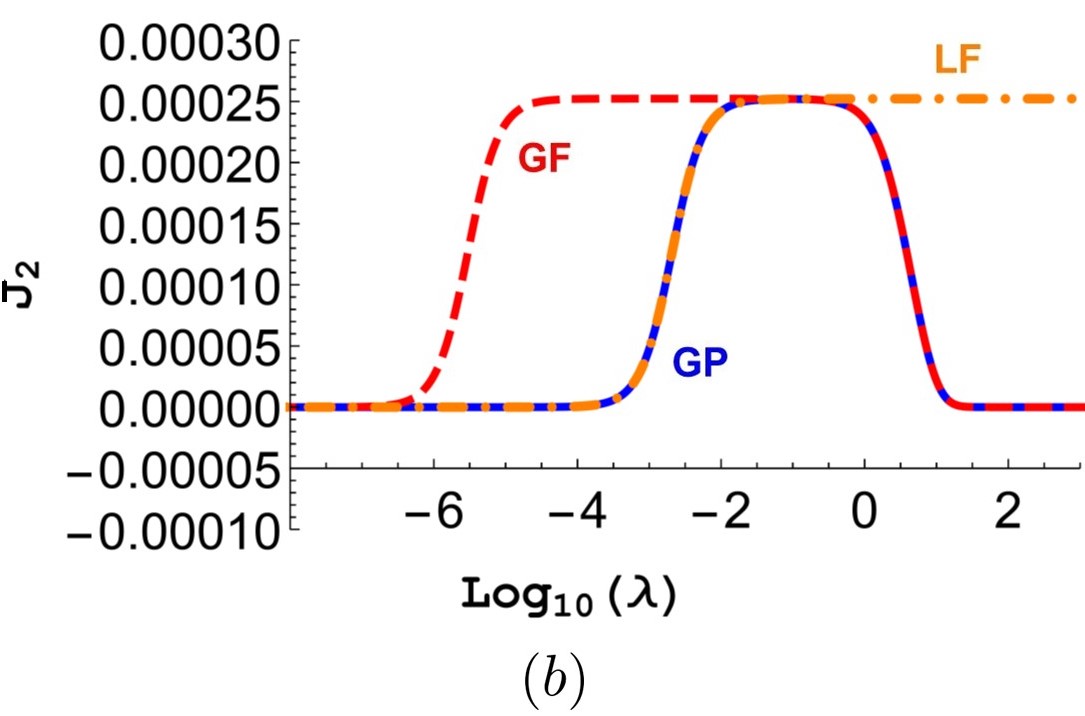}
\caption{Incoming heat current from the hot reservoir as a function of the coupling constant, in the presence of two separate baths with inverse temperatures $\beta^{(l_1)}=1$, $\beta^{(l_2)}=0.1$, for $\omega_-=10^{-3}$ (a) and $\omega_-=10^{-5}$ (b). We confront the value obtained through the global master equation (m.e.) with partial secular approximation GP (solid blue), the global m.e. with full
secular approximation GF (dashed red), and the local m.e. with full secular
approximation LF (dot-dashed orange), which coincides with the local m.e. with partial secular approximation. When $\lambda\ll 1$ both the global m.e. with partial secular approximation and the local master equation correctly reproduce the value of the heat current, which decreases toward zero for $\lambda\rightarrow 0$. On the contrary, the global m.e. with full secular approximation highly overestimates the value of the current when the detuning is not way higher than the coupling constant. Starting from $\lambda=O(1)$, the local approach incorrectly produces a stationary non-zero value of the heat current, while the global approach accurately describes the decay of the current toward zero.}
\label{fig:current}
\end{figure}

Fig.~\ref{fig:current} depicts the stationary heat current incoming from the hot bath as a function of the coupling constant $\lambda$, with detuning $\omega_-=10^{-3}$ (a) and $\omega_-=10^{-5}$ (b), for distinct master equations. We see that there is a region of $\lambda$ in which the global master equation with full secular approximation fails, highly overestimating the value of the current: this is the effect that was extensively observed and discussed in the recent works on the topic \cite{Gonzalez2017,Hofer2017}. The full secular approximation breaks down because $\omega_-\ll 1$ and $\lambda\ll 1$, making small the energy difference in equation~\ref{eqn:energyDiff}. However, we also observe that such region depends on the ratio $\omega_-/\lambda$, as discussed in Sec.~\ref{sec:comparison} and displayed in Table~\ref{tab:tabella}: if $\omega_-\gg \lambda$, the eigenmodes basis of the system is almost local and the global approach with full secular is accurate as well. For this reason, the range in which 
it fails is more narrow for $\omega_-=10^{-3}$ than for $\omega_-=10^{-5}$. If the detuning were null, the global approach with full secular approximation would provide a non-zero heat current even for $\lambda\rightarrow 0$ \cite{Hofer2017}, which is clearly non-physical. On the contrary, the global approach with partial secular approximation and the local approach (for which partial and full secular coincide) provide a correct description of the 
incoming heat current for small values of $\lambda$. Increasing the coupling constant, we observe that the current increases as well till reaching the value given by the full secular approximation, which starting from here recovers its validity. For big values of $\lambda$ the global master equations describe a current decreasing toward $0$. This is correct, since if $\lambda\gg 1$ the only relevant part of the system Hamiltonian $H_S$ is $\lambda\sigma_1^x\sigma_2^x$, and therefore it commutes with the interaction Hamiltonian: $\lim_{\lambda\rightarrow \infty}[H_S,
H_I]=0$. Hence, the dissipator associated to each bath does not induce an energy exchange in 
any stationary state (which now depends on the initial conditions), and no heat current is 
produced. On the contrary, the local master equation is written in a basis which is different 
from the diagonal basis of $H_S$, and thus indicates a fictitious non-zero heat current even 
for $\lambda\gg 1$. Note that the global approach with partial secular approximation 
 is appropriate in all considered parameter regimes.

The reader can verify that the failure of the local master equation or of the full secular approximation to describe the incoming heat current in different regimes correctly reproduces the parameters ranges summarized in Table~\ref{tab:tabella}.

\section{Concluding remarks}
\label{sec:Conclusions}
\begin{table}[h]
\begin{center}
\begin{tabular}{cccc}
\multicolumn{4}{c}{\textbf{Validity of the master equation}}\\[4.5pt]
\toprule
&&Common bath&Separate baths\\
\midrule
\multirow{2}*{Global m.e.}&Partial secular&always&always \\ \cmidrule{2-4}
                   &Full secular&$\lambda\gg\mu^2$ or 
$\omega_-\gg\mu^2$&$\lambda\gg\mu^2$ or $\omega_-\gg\lambda$  \\
                   \midrule
\multirow{2}*{Local m.e.}&Partial secular&$\lambda\ll 1$ & $\lambda\ll 1$ \\ 
\cmidrule{2-4}
                   &Full secular&$\lambda\ll 1$ and $\omega_-\gg \mu^2$ & 
$\lambda\ll 1$ \\
\bottomrule
\end{tabular}
\end{center}
\caption{Conditions of validity of each master equation (m.e.), depending on the 
values of the detuning between the qubits $\omega_-$, of the qubit-qubit 
coupling constant $\lambda$ and of the qubit-bath coupling constant $\mu$. Each 
possible scenario is taken into account: local or global master equation, 
partial or full secular approximation, presence of a common as well as separate 
baths. We recall that all the constants are dimensionless quantities, according 
to the renormalization discussed in 
equations~\ref{eqn:freeSystemHamInit},~\ref{eqn:freeSystemHamInitRen} 
and~\ref{eqn:HamiltonianGeneral}, and that the general condition for the 
validity of the secular approximation is presented in 
equation~\ref{eqn:secularAppCond}. The table only deals with scenarios in which the Born-Markov approximations hold. Furthermore, we recall that the local approach is always valid in the singular-coupling scenario, independently of the system parameters.}
\label{tab:tabella}
\end{table}

In the present work we have extensively addressed the derivation of the 
Markovian master equation for two qubits interacting with thermal baths,
in order to assess the validity of local and global master equations.  
A comprehensive description is achieved considering all the possible 
scenarios: presence of separate as well as a common bath, including both 
dissipative and dephasing interaction, and taking into account the possibility 
of a direct coupling between the qubits and of detuning. 
The Markovian master equation, in the form of a Bloch-Redfield master equation, 
has been derived in Sec.~\ref{sec:Derivation} reviewing all the necessary 
approximations, carefully stating the condition for the validity of the Born and the Markov approximations.  We have 
obtained two general master equations: equation~\ref{eqn:masterEqNoDir} in the case 
without a direct qubit-qubit coupling, and equation~\ref{eqn:masterEqDir} when a 
direct coupling between the qubits must be taken into account.

These preliminary steps put ourselves in the conditions of determining the 
validity of the final possible approximation, namely the secular one. We have 
established the requirements under which one can apply a \textit{full secular} or 
only a \textit{partial secular} approximation. This assessment is especially 
relevant in the context of the feasibility of a \textit{local approach} to the 
master equation with respect to the \textit{global} one, which may simplify the 
computation of the solution in many cases. This renewed problem is addressed in  
 Sec.~\ref{sec:localGlobal}.

Many works have already proven that the global master equation may fail in some scenarios because of the breaking of the (full) secular approximation. In this paper, we have shown how to overcome such problem by applying an accurate \textit{partial} secular approximation which does not remove slowly-rotating terms. In particular, in Sec.~\ref{sec:Derivation} we have provided an extensive mathematical derivation of such equation, termed global master equation with partial 
secular approximation, proving that it is \textit{always} the most correct choice in any parameters scenario in which 
the Born and Markov approximations are valid. Then, in Sec.~\ref{sec:localGlobal} we have shown how to derive the local master equation, and focused on the comparison between it and the global approach for the case of two coupled qubits, with partial or full secular approximation. The local master equation is always accurate in the singular-coupling limit, as already extensively proven \cite{BreuerPetruccione,Hofer2017}. If this limit is not assumed, the local approach is valid 
only for small values of the qubit-qubit coupling constant $\lambda\ll 1$. The 
feasibility of the full secular approximation must be checked: if a common bath 
is present, the detuning between the qubits plays a fundamental role, since a 
small value of it would make the local master equation with full secular 
approximation fail; in the case of the global master equation, the value of 
$\lambda$ comes into play as well. If the qubits interact with separate baths 
only, some 
subtleties emerge: for the local master equation the full secular approximation 
is always valid, while in the global approach care must be taken, since if both 
$\lambda$ and the detuning are very small, the condition for the approximation 
may break down. We have shown that 
the value of the qubit-qubit coupling constant $\lambda$ is not the only important 
actor here: the ratio between $\lambda$ and the detuning $\omega_-$ must be 
considered as well, since if $\omega_-\gg\lambda$ the full secular approximation 
in the global approach recovers its validity. These results are summarized in
Table~\ref{tab:tabella}, where we have highlighted the scenarios with local or 
global approach, partial or full secular approximation, and common or separate 
baths.
At the end of Sec.~\ref{sec:localGlobal}, we have discussed the consequences of the 
local vs global issue in the case of two qubits, and we have compared the 
results of the system evolution obtained through different approaches by depicting 
them in figures~\ref{fig:NoDet},~\ref{fig:steady} and~\ref{fig:ComPar}. 

In Sec.~\ref{sec:physicalQ} we have shown how several physical quantities change when been computed through different master equations. In particular, in the case of weak qubit-qubit coupling constant, small detuning and a common bath, the full secular approximation is not able to detect important phenomena such as quantum beats and quantum synchronization, and does not produce entanglement during the evolution as depicted in Fig.~\ref{fig:nega}. In the scenario with two separate baths with unbalanced temperatures, the global master equation with full secular approximation leads to an unphysical stationary heat current in the region in which the detuning $\omega_-$ is not way larger than $\lambda$. On the contrary, after a transient in which the stationary heat current is correctly detected, it is the local master equation which predicts a non-zero fictitious current when $\lambda\gg 1$. The global master equation with partial secular approximation reproduces the correct physical results in any scenario 
and any range of parameters.

Our discussion remains valid while considering general interacting bipartite systems and can also be 
immediately extended to multipartitite scenarios, while more challenging will be to explore local 
versus global approaches in non-Markovian situations.
Also, this analysis is relevant for  extended systems experiencing dissipation only in 
some of their parts, so that global 
master equations need in principle to be considered \cite{giorgi2016}.
Our general conclusions are especially timely, because of the renewed interest on the topic that has lead to several results, sometimes contradictory or only partial. 
Beyond being a fundamental instrument for the appropriate description of 
coupled qubits in contact with environments,
further implications may be foreseen in the context of thermodynamics, computations and information, 
considering the differences arising between  phenomenological approaches and the microscopically derived 
global master equation with partial secular approximation.

\section*{Acknowledgments}

The authors acknowledge funding from MINECO/AEI/FEDER through projects EPheQuCS
FIS2016-78010-P, the María de Maeztu Program
for Units of Excellence in R$\&$D (MDM-2017-0711), and the
CAIB postdoctoral program, and support from CSIC Research Platform PTI-001.
M.C. acknowledges partial funding from Fondazione Angelo della Riccia.

\appendix
\section{Generality of the system Hamiltonian}
\label{sec:generalitySystem}
\subsection{Qubit rotations }
\label{sec:Rotat}
Following the standard convention in quantum information theory, we have chosen 
to write throughout all the paper the free Hamiltonian of a single qubit as 
$H_1=\omega_1/2 \sigma_1^z$. Historically, this has not been always the 
preferred choice, since for instance in the seminal papers by Caldeira and 
Leggett \cite{Leggett1987} a different notation was employed, in particular:
\begin{equation}
\label{eqn:CaldeiraLeggett}
H_1'=\frac{\epsilon_1}{2}\sigma_1^z+\frac{\Delta_1}{2}\sigma_1^x,
\end{equation}
which regards the two level system as two levels with detuning $\epsilon$ 
separated by a potential barrier, with the possibility of hopping through it via 
$\sigma_1^x$. 

$H_1$ and $H_1'$ are connected by a unitary transformation that, in absence of interactions between the qubits,  will 
change the dephasing and dissipative character of each bath, and in this work both possibilities are taken into account.
On the other hand, in presence of the coupling between qubits, the local unitary transformations will in general transform the 
 coupling  $H_{12}=\lambda\sigma_1^x\sigma_2^x$ into a more complex interaction term.
 
\subsection{Further qubit-qubit couplings}
\label{sec:couplingsQQ}
In this section we address further possible forms of the qubit-qubit 
interaction. We remind the system Hamiltonian:
\begin{equation}
\label{eqn:systemHamBis}
H_S=\frac{\omega_1}{2}\sigma_1^z+\frac{\omega_2}{2}\sigma_2^z+H_{12},
\end{equation}
with  $H_{12}=\lambda\sigma_1^x\sigma_2^x$. This choice is 
justified by the fact that such interaction is the standard one employed in the 
framework of quantum information, to which this paper is mostly devoted; we 
indeed find the $\sigma_1^x\sigma_2^x$ interaction (or fully equivalently 
$\sigma_1^y\sigma_2^y$) in many experimental platforms, such as superconducting 
qubits \cite{Majer2007a,Mastellone2008,Wendin2007a} or coupled atomic dipoles 
\cite{Stokes2018}. Anyway, let us examine possible alternatives and the connection with this Ising-like coupling.

\subsubsection{Heisenberg-type interaction}
Let's here consider two qubits coupled
through an Heisenberg-type interaction, i.e. 
$H_{12}=\sum_{k=x,y,z}\lambda_k\sigma_1^k\sigma_2^k$, quite common in many physical systems. This interaction conserves the 
\textit{parity} of the number of excitations and, for this reason, the 
system Hamiltonian in the canonical basis has the same structure of 
equation~\ref{eqn:systemHamDir}:
\begin{equation}
\label{eqn:systemHamHeisenberg}
H_S=\left(\begin{array}{cccc}
\omega_+/2+\lambda_z&0&0&\lambda_x-\lambda_y\\
0&\omega_-/2-\lambda_z&\lambda_x+\lambda_y&0\\
0&\lambda_x+\lambda_y&-\omega_-/2-\lambda_z&0\\
\lambda_x-\lambda_y&0&0&-\omega_+/2+\lambda_z
\end{array}\right),
\end{equation}
with $\omega_\pm=\omega_1\pm\omega_2$.

The parity symmetry makes the eigenvectors remain of the same form of the 
eigenvectors in equation~\ref{eqn:eigenVecDir}. The only thing that may change is the 
value of the angles $\theta$ and $\phi$ in equation~\ref{eqn:paramTheta}, according 
to the different values of $\lambda_k$ appearing in 
equation~\ref{eqn:systemHamHeisenberg}. This means that the general master equation 
preserves the structure in equation~\ref{eqn:masterEqDir}: the value of the 
coefficients marks the only difference between Heisenberg- and Ising-type 
interaction.

\subsubsection{Rotating wave approximation}
Another very common case in the 
literature is to consider the qubit-qubit coupling in RWA, i.e.
$H_{12}=\lambda(\sigma_1^-\sigma_2^++\sigma_1^+\sigma_2^-)$. Even if in some cases this can be justified,
we notice that without counter-rotating terms the Hamiltonian not only conserves 
the parity of the number of excitations, but also conserves the number of excitations 
itself. A standard rotation is sufficient to diagonalize it, leading to the eigenvalues:
\begin{equation}
\label{eqn:eigenValRW}
\begin{aligned}
E_3&=+\omega_+/2,\\
E_2&=+\sqrt{\lambda^2+\omega_-^2/4},\\
E_1&=-\sqrt{\lambda^2+\omega_-^2/4},\\
E_0&=-\omega_+/2,
\end{aligned}
\end{equation}
with associated eigenvectors
\begin{equation}
\label{eqn:eigenVecRW}
\begin{aligned}
\ket{e_3}&=\ket{11},\\
\ket{e_2}&=\cos\phi\ket{10}+\sin\phi\ket{01},\\
\ket{e_1}&=\sin\phi\ket{10}-\cos\phi\ket{01},\\
\ket{e_0}&=\ket{00},
\end{aligned}
\end{equation}
where the angle $\phi$ is the same as in equation~\ref{eqn:paramTheta}:
\begin{equation}
\label{eqn:phiAngle}
\sin 2\phi=\frac{\lambda}{E_2},\qquad\cos 2\phi=\frac{\omega_-/2}{E_2}.
\end{equation}

We see that, unless we can neglect the coupling ($\lambda\ll\omega_+$), the geometry of the 
Hamiltonian in the RWA is non-trivially different from the one used in the work. In the master equation, the absence of 
counter-rotating terms eliminates all the ``double emission'' or ``double absorption'' jump 
operators, namely $\sigma_j^z(\pm\omega_{III})$ in equation~\ref{eqn:jumpOpDir}. 
Without these terms important effects do not arise, such as  stationary entanglement \cite{Scala2011},
or are not properly described, as the refrigerator  performance analyzed in Ref. \cite{seah}.

\section{Coefficients of the master equation}
\label{sec:coeff}
\subsection{No direct coupling}
The coefficients of the master equation in equation~\ref{eqn:masterEqNoDir} read:
\begin{eqnarray}
\label{eqn:coeffNoDir}
\gamma_{jk}=\left\{\begin{array}{lr}
(g_x^{(c_1)}g_x^{(c_2)})\left[\Gamma^{(c)}(\omega_j)+(\Gamma^{(c)}
(\omega_k))^*\right]&\text{if } j\neq k,\\
\sum_{\alpha=l_j,c_j} 
(g_x^{(\alpha)})^2\left[\Gamma^{(\alpha)}(\omega_j)+(\Gamma^{(\alpha)}
(\omega_j))^* \right]&\text{if } j=k,\\
\end{array}\right.\nonumber \\
\tilde{\gamma}_{jk}=\left\{\begin{array}{lr}
(g_x^{(c_1)}g_x^{(c_2)})\left[\Gamma^{(c)}(-\omega_j)+(\Gamma^{(c)}
(-\omega_k))^*\right]&\text{if } j\neq k,\\
\sum_{\alpha=l_j,c_j} 
(g_x^{(\alpha)})^2\left[\Gamma^{(\alpha)}(-\omega_j)+(\Gamma^{(\alpha)}
(-\omega_j))^* \right]&\text{if } j=k,\\
\end{array}\right.\nonumber \\
\eta_{jk}=\left\{\begin{array}{lr}
(g_z^{(c_1)}g_z^{(c_2)})\left[\Gamma^{(c)}(0)+(\Gamma^{(c)}(0))^*\right]&\text{
if } j\neq k,\\
\sum_{\alpha=l_j,c_j} 
(g_z^{(\alpha)})^2\left[\Gamma^{(\alpha)}(0)+(\Gamma^{(\alpha)}(0))^* 
\right]&\text{if } j=k,\\
\end{array}\right. \\
s_{jk}=\left\{\begin{array}{lr}
(g_x^{(c_1)}g_x^{(c_2)})\left[\Gamma^{(c)}(\omega_j)-(\Gamma^{(c)}
(\omega_k))^*\right]/2i&\text{if } j\neq k,\\
\sum_{\alpha=l_j,c_j} 
(g_x^{(\alpha)})^2\left[\Gamma^{(\alpha)}(\omega_j)-(\Gamma^{(\alpha)}
(\omega_j))^*\right]/2i&\text{if } j=k,\\
\end{array}\right.\nonumber\\
\tilde{s}_{jk}=\left\{\begin{array}{lr}
(g_x^{(c_1)}g_x^{(c_2)})\left[\Gamma^{(c)}(-\omega_j)-(\Gamma^{(c)}
(-\omega_k))^*\right]/2i&\text{if } j\neq k,\\
\sum_{\alpha=l_j,c_j} 
(g_x^{(\alpha)})^2\left[\Gamma^{(\alpha)}(-\omega_j)-(\Gamma^{(\alpha)}
(-\omega_j))^*\right]/2i&\text{if } j=k,\\
\end{array}\right.\nonumber \\
s_0=(g_z^{(c_1)}g_z^{(c_2)})\frac{\Gamma^{(c)}(0)-(\Gamma^{(c)}(0))^*}{2i},
\qquad\qquad\qquad\! s_1=s_{11}-\tilde{s}_{11},\nonumber \\
s_2=s_{22}-\tilde{s}_{22},\qquad\qquad s_+=s_{12}+\tilde{s}_{21},\qquad\qquad 
s_-=s_{21}+\tilde{s}_{12}.\nonumber
\end{eqnarray}
$\Gamma^{(\alpha)}(\omega)$ are the one-side Fourier transforms of the baths 
correlation functions defined in equation~\ref{eqn:oneSideBath}, where with abuse of 
notation $\alpha=c_1=c_2=c$. The mean value is performed on the thermal state 
$\rho_B^{(\alpha)}$ of each bath, at a given inverse temperature 
$\beta^{(\alpha)}$~\cite{BreuerPetruccione}. $g_j^{(\alpha)}$ are the coupling 
constants expressing the strength of the qubit-bath interactions. In the 
Lamb-Shift Hamiltonian equation~\ref{eqn:lambShiftNoDir}, $s_1/2\sigma_1^z$ and 
$s_2/2\sigma_2^z$ lead to a renormalization of the qubit frequencies. Moreover, 
note that we have neglected any multiple of the identity appearing in the 
Lamb-Shift Hamiltonian.

\subsection{Direct coupling}
Analogously to equation~\ref{eqn:coeffNoDir}, we list the coefficients appearing in 
the master equation in equation~\ref{eqn:masterEqDir}:
\begin{eqnarray}
\label{eqn:coefficientsMasterEqDir}
\gamma_{jk}^{mn}=\left\{\begin{array}{lr}
(g_x^{(c_1)}g_x^{(c_2)})\left[\Gamma^{(c)}(\omega_j)+(\Gamma^{(c)}
(\omega_k))^*\right]&\text{if } m\neq n,\\
\sum_{\alpha=l_m,c_m} 
(g_x^{(\alpha)})^2\left[\Gamma^{(\alpha)}(\omega_j)+(\Gamma^{(\alpha)}
(\omega_k))^* \right]&\text{if } m=n,\\
\end{array}\right.\nonumber\\
\tilde{\gamma}_{jk}^{mn}=\left\{\begin{array}{lr}
(g_x^{(c_1)}g_x^{(c_2)})\left[\Gamma^{(c)}(-\omega_j)+(\Gamma^{(c)}
(-\omega_k))^*\right]&\text{if } m\neq n,\\
\sum_{\alpha=l_m,c_m} 
(g_x^{(\alpha)})^2\left[\Gamma^{(\alpha)}(-\omega_j)+(\Gamma^{(\alpha)}
(-\omega_k))^* \right]&\text{if } m=n,\\
\end{array}\right.\nonumber\\
\eta_{jk}^{mn}=\left\{\begin{array}{lr}
(g_z^{(c_1)}g_z^{(c_2)})\left[\Gamma^{(c)}(\omega_j)+(\Gamma^{(c)}
(\omega_k))^*\right]&\text{if } m\neq n,\\
\sum_{\alpha=l_m,c_m} 
(g_z^{(\alpha)})^2\left[\Gamma^{(\alpha)}(\omega_j)+(\Gamma^{(\alpha)}
(\omega_k))^* \right]&\text{if } m=n,\\
\end{array}\right.\nonumber\\
\zeta_{j}^{mn}=\left\{\begin{array}{lr}
(g_z^{(c_1)}g_z^{(c_2)})\left[\Gamma^{(c)}(\omega_j)+(\Gamma^{(c)}
(\omega_j))^*\right]&\text{if } m\neq n,\\
\sum_{\alpha=l_m,c_m} 
(g_z^{(\alpha)})^2\left[\Gamma^{(\alpha)}(\omega_j)+(\Gamma^{(\alpha)}
(\omega_j))^* \right]&\text{if } m=n,\\
\end{array}\right.\\
s_{jk}^{mn}=\left\{\begin{array}{lr}
(g_x^{(c_1)}g_x^{(c_2)})\left[\Gamma^{(c)}(\omega_j)-(\Gamma^{(c)}
(\omega_k))^*\right]/2i &\text{if } m\neq n,\\
\sum_{\alpha=l_m,c_m} 
(g_x^{(\alpha)})^2\left[\Gamma^{(\alpha)}(\omega_j)-(\Gamma^{(\alpha)}
(\omega_k))^* \right]/2i &\text{if } m=n,\\
\end{array}\right.\nonumber\\
\tilde{s}_{jk}^{mn}=\left\{\begin{array}{lr}
(g_x^{(c_1)}g_x^{(c_2)})\left[\Gamma^{(c)}(-\omega_j)-(\Gamma^{(c)}
(-\omega_k))^*\right]/2i &\text{if } m\neq n,\\
\sum_{\alpha=l_m,c_m} 
(g_x^{(\alpha)})^2\left[\Gamma^{(\alpha)}(-\omega_j)-(\Gamma^{(\alpha)}
(-\omega_k))^* \right]/2i&\text{if } m=n,\\
\end{array}\right.\nonumber\\
r_{jk}^{mn}=\left\{\begin{array}{lr}
(g_z^{(c_1)}g_z^{(c_2)})\left[\Gamma^{(c)}(\omega_j)-(\Gamma^{(c)}
(\omega_k))^*\right]/2i &\text{if } m\neq n,\\
\sum_{\alpha=l_m,c_m} 
(g_z^{(\alpha)})^2\left[\Gamma^{(\alpha)}(\omega_j)-(\Gamma^{(\alpha)}
(\omega_k))^* \right]/2i &\text{if } m=n,\\
\end{array}\right.\nonumber\\
u_{j}^{mn}=\left\{\begin{array}{lr}
(g_z^{(c_1)}g_z^{(c_2)})\left[\Gamma^{(c)}(\omega_j)-(\Gamma^{(c)}
(\omega_j))^*\right]/2i &\text{if } m\neq n,\\
\sum_{\alpha=l_m,c_m} 
(g_z^{(\alpha)})^2\left[\Gamma^{(\alpha)}(\omega_j)-(\Gamma^{(\alpha)}
(\omega_j))^* \right]/2i &\text{if } m=n.\\
\end{array}\right.\nonumber
\end{eqnarray}

\section{Jump operators of the Hamiltonian with direct coupling}
\label{sec:jump}
\begin{figure}
\center
\includegraphics[scale=0.35]{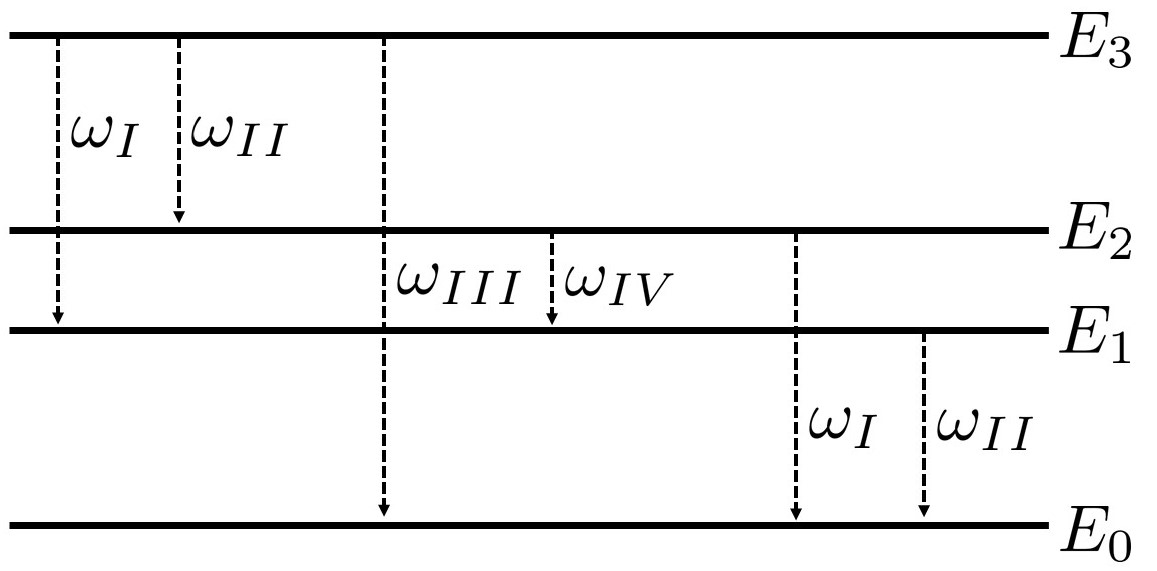}
\caption{Diagram of the states of the system Hamiltonian 
equation~\ref{eqn:systemHamDir}, with all the possible emission frequencies.}
\label{fig:schemeDir}
\end{figure}

To find the jump operators we first of all have to recognize which are the jump 
frequencies associated to each system operator. Recalling the eigenvalues in 
equation~\ref{eqn:eigenValDir}, we can recognize four different frequencies (positive 
or negative) of the jumps between different eigenvalues, in addition to the zero 
frequency, which are depicted in figure~\ref{fig:schemeDir}. We name them as:
\begin{equation}
\label{eqn:jumpFrequencies}
\begin{aligned}
&\omega_I=E_3-E_1=E_2-E_0=\sqrt{\lambda^2+\omega_+^2/4}+\sqrt{
\lambda^2+\omega_-^2/4},\\
&\omega_{II}=E_3-E_2=E_1-E_0=\sqrt{\lambda^2+\omega_+^2/4}-\sqrt{
\lambda^2+\omega_-^2/4},\\
&\omega_{III}=E_3-E_0=2\sqrt{\lambda^2+\omega_+^2/4},\\
&\omega_{IV}=E_2-E_1=2\sqrt{\lambda^2+\omega_-^2/4}.\\
\end{aligned}
\end{equation}

By employing the definition in equation~\ref{eqn:jumpOp}, we can easily see that the 
operators $\sigma_j^x$ with $j=1,2$ only induce transitions of frequency 
$\omega_I$ and $\omega_{II}$, while $\sigma_j^z$ are responsible for the 
excitations and decaying of frequency $\omega_{III}$, $\omega_{IV}$ and $0$.

In the following we list all the jump operators, where we recall that we are 
using the notation of equation~\ref{eqn:jumpOp} for the jump operators; for instance, 
$\sigma_1^x(\omega_I)=\sum_{\epsilon'-\epsilon=\omega_I}\ket{\epsilon}\bra{
\epsilon}{\sigma_1^x}\ket{\epsilon'}\bra{\epsilon'}$:
\begin{equation}
\label{eqn:jumpOpDir}
\begin{aligned}
&\sigma_1^x(\omega_I)=\cos(\theta+\phi)(\ket{e_0}\bra{e_2}+\ket{e_1}\bra{e_3}),
\\
&\sigma_1^x(\omega_{II})=\sin(\theta+\phi)(-\ket{e_0}\bra{e_1}+\ket{e_2}\bra{e_3
}),\\
&\sigma_2^x(\omega_{I})=\sin(\theta-\phi)(-\ket{e_0}\bra{e_2}+\ket{e_1}\bra{e_3}
),\\
&\sigma_2^x(\omega_{II})=\cos(\theta-\phi)(\ket{e_0}\bra{e_1}+\ket{e_2}\bra{e_3}
),\\
&\sigma_1^z(0)=\cos 2\theta(\ket{e_3}\bra{e_3}-\ket{e_0}\bra{e_0})+\cos 
2\phi(\ket{e_2}\bra{e_2}-\ket{e_1}\bra{e_1}),\\
&\sigma_1^z(\omega_{III})=-\sin 2\theta(\ket{e_0}\bra{e_3}),\\
&\sigma_1^z(\omega_{IV})=-\sin 2\phi(\ket{e_1}\bra{e_2}),\\
&\sigma_2^z(0)=\cos 2\theta(\ket{e_3}\bra{e_3}-\ket{e_0}\bra{e_0})+\cos 
2\phi(\ket{e_1}\bra{e_1}-\ket{e_2}\bra{e_2}),\\
&\sigma_2^z(\omega_{III})=-\sin 2\theta(\ket{e_0}\bra{e_3}),\\
&\sigma_2^z(\omega_{IV})=+\sin 2\phi(\ket{e_1}\bra{e_2}).\\
\end{aligned}
\end{equation}
The jump operators with negative frequencies are obtained by employing the 
property in equation~\ref{eqn:propJumpOp} $A_\beta(-\omega)=A_\beta(\omega)^\dagger$.

Notice that, once again, the jump operators associated to $\sigma_j^x$ and 
$\sigma_j^z$ have different frequencies, whose difference is not ``small'' in 
the sense of the condition for the secular approximation 
equation~\ref{eqn:secularAppCond}. Actually, there may be a singular case in 
which two frequencies of different bath operators assume the same value, namely 
$\omega_{II}$ and $\omega_{IV}$:
\begin{equation}
\label{eqn:frequencyCrossing}
\omega_{II}-\omega_{IV}=\sqrt{\lambda^2+\omega_+^2/4}-3\sqrt{
\lambda^2+\omega_-^2/4}.
\end{equation}
By setting the above equation equal to zero, we find the condition for which we 
must consider the ``crossing'' between $\omega_{II}$ and $\omega_{IV}$ in the 
master equation, i.e. the values of the constants for which we cannot 
neglect these cross terms in the master equation. The condition reads: 
\begin{equation}
\label{eqn:conditionCrossing}
32\lambda^2=\omega_+^2-9\omega_-^2.
\end{equation}
Anyway, we can see that this case is only ``singular'', in the sense that it 
regards a ``zero-measure'' region in the parameter space. Indeed, even if the 
values of $\lambda,\omega_1,\omega_2$ satisfy equation~\ref{eqn:conditionCrossing}, 
it is sufficient to perturb one of them by a quantity of order larger than 
$O(\mu^2)$ to be allowed to neglect the cross terms, since they would fulfil the 
condition in equation~\ref{eqn:secularAppCond}.
Therefore, in this paper we do not discuss the singular case in which we 
need to conserve cross terms between $\sigma^x$ and $\sigma^z$, since this 
would tangibly complicate the master equation. Then, the argument about the separation of the dephasing and dissipative bath 
discussed in Sec.~\ref{sec:noDirect} holds. 

\bibliographystyle{iopart-num}      
\bibliography{Report}

\providecommand{\newblock}{}
\begin{thebibliography}{10}
\expandafter\ifx\csname url\endcsname\relax
  \def\url#1{{\tt #1}}\fi
\expandafter\ifx\csname urlprefix\endcsname\relax\def\urlprefix{URL }\fi
\providecommand{\eprint}[2][]{\url{#2}}

\bibitem{Storcz2003}
Storcz M~J and Wilhelm F~K 2003 {\em Phys. Rev. A\/} {\bf 67} 042319

\bibitem{PhysRevLett.89.147902}
Levy J 2002 {\em Phys. Rev. Lett.\/} {\bf 89} 147902

\bibitem{Haddadfarshi_2016}
Haddadfarshi F and Mintert F 2016 {\em New J. Phys.\/} {\bf 18} 123007

\bibitem{Reiter2017}
Reiter F, S{\o}rensen A~S, Zoller P and Muschik C~A 2017 {\em Nat. Commun.\/}
  {\bf 8} 1822

\bibitem{PhysRevX.7.041009}
Demkowicz-Dobrza\ifmmode~\acute{n}\else \'{n}\fi{}ski R, Czajkowski J and
  Sekatski P 2017 {\em Phys. Rev. X\/} {\bf 7} 041009

\bibitem{PhysRevLett.89.277901}
Braun D 2002 {\em Phys. Rev. Lett.\/} {\bf 89} 277901

\bibitem{Benatti2003a}
Benatti F, Floreanini R and Piani M 2003 {\em Phys. Rev. Lett.\/} {\bf 91}
  070402

\bibitem{Horn_2018}
Horn K~P, Reiter F, Lin Y, Leibfried D and Koch C~P 2018 {\em New J. Phys.\/}
  {\bf 20} 123010

\bibitem{Barreiro2011}
Barreiro J~T, M{\"u}ller M, Schindler P, Nigg D, Monz T, Chwalla M, Hennrich M,
  Roos C~F, Zoller P and Blatt R 2011 {\em Nature\/} {\bf 470} 486–491

\bibitem{Lin2013}
Lin Y, Gaebler J~P, Reiter F, Tan T~R, Bowler R, Sørensen A~S, Leibfried D and
  Wineland D~J 2013 {\em Nature\/} {\bf 504} 415–418

\bibitem{PhysRevLett.116.240503}
Kimchi-Schwartz M~E, Martin L, Flurin E, Aron C, Kulkarni M, Tureci H~E and
  Siddiqi I 2016 {\em Phys. Rev. Lett.\/} {\bf 116} 240503

\bibitem{PhysRevX.7.011016}
Fitzpatrick M, Sundaresan N~M, Li A~C~Y, Koch J and Houck A~A 2017 {\em Phys.
  Rev. X\/} {\bf 7} 011016

\bibitem{Linden2010}
Linden N, Popescu S and Skrzypczyk P 2010 {\em Phys. Rev. Lett.\/} {\bf 105}
  130401

\bibitem{Bohr_Brask_2015}
Brask J~B, Haack G, Brunner N and Huber M 2015 {\em New J. Phys.\/} {\bf 17}
  113029

\bibitem{PhysRevE.99.042135}
Manzano G, Silva R and Parrondo J~M~R 2019 {\em Phys. Rev. E\/} {\bf 99} 042135

\bibitem{Dicke1954}
Dicke R~H 1954 {\em Phys. Rev.\/} {\bf 93} 99--110

\bibitem{Giorgi2013}
Giorgi G~L, Plastina F, Francica G and Zambrini R 2013 {\em Phys. Rev. A\/}
  {\bf 88} 042115

\bibitem{Ficek2002}
Ficek Z and Tana{\'{s}} R 2002 {\em Phys. Rep.\/} {\bf 372} 369--443

\bibitem{Scala2008}
Scala M, Migliore R and Messina A 2008 {\em J. Phys. A Math. Theor.\/} {\bf 41}
  435304

\bibitem{Li2009a}
Li J and Paraoanu G~S 2009 {\em New J. Phys.\/} {\bf 11} 113020

\bibitem{Campagnano2010}
Campagnano G, Hamma A and Weiss U 2010 {\em Phys. Lett. A\/} {\bf 374} 416--423

\bibitem{Orth2010}
Orth P~P, Roosen D, Hofstetter W and {Le Hur} K 2010 {\em Phys. Rev. B\/} {\bf
  82} 144423

\bibitem{Scala2011}
Scala M, Migliore R, Messina A and S{\'{a}}nchez-Soto L~L 2011 {\em Eur. Phys.
  J. D\/} {\bf 61} 199--205

\bibitem{Santos2014}
Santos J~P and Semi{\~{a}}o F~L 2014 {\em Phys. Rev. A\/} {\bf 89} 022128

\bibitem{Levy2014}
Levy A and Kosloff R 2014 {\em Europhys. Lett.\/} {\bf 107} 20004

\bibitem{Trushechkin2016}
Trushechkin A~S and Volovich I~V 2016 {\em Europhys. Lett.\/} {\bf 113} 30005

\bibitem{Gonzalez2017}
Gonz{\'{a}}lez J~O, Correa L~A, Nocerino G, Palao J~P, Alonso D and Adesso G
  2017 {\em Open Syst. Inf. Dyn.\/} {\bf 24} 1740010

\bibitem{Hofer2017}
Hofer P~P, Perarnau-Llobet M, Miranda L~D~M, Haack G, Silva R, Brask J~B and
  Brunner N 2017 {\em New J. Phys.\/} {\bf 19} 123037

\bibitem{Mitchison2018}
Mitchison M~T and Plenio M~B 2018 {\em New J. Phys.\/} {\bf 20} 033005

\bibitem{Koch2007a}
Koch J, Yu T~M, Gambetta J, Houck A~A, Schuster D~I, Majer J, Blais A, Devoret
  M~H, Girvin S~M and Schoelkopf R~J 2007 {\em Phys. Rev. A\/} {\bf 76} 042319

\bibitem{BreuerPetruccione}
Breuer H~P and Petruccione F 2002 {\em The theory of open quantum systems\/}
  (Oxford University Press)

\bibitem{Nakajima1958}
Nakajima S 1958 {\em Prog. Theor. Phys.\/} {\bf 20} 948--959

\bibitem{Zwanzig1960}
Zwanzig R 1960 {\em J. Chem. Phys.\/} {\bf 33} 1338--1341

\bibitem{Rivas2010a}
Rivas {\'{A}}, {K Plato} A~D, Huelga S~F and Plenio M~B 2010 {\em New J.
  Phys.\/} {\bf 12} 113032

\bibitem{Cresser2017a}
Cresser J and Facer C 2017 {\em preprint arXiv:1710.09939\/}

\bibitem{Farina2019}
Farina D and Giovannetti V 2019 {\em Phys. Rev. A\/} {\bf 100} 012107

\bibitem{Albash2012}
Albash T, Boixo S, Lidar D~A and Zanardi P 2012 {\em New J. Phys.\/} {\bf 14}
  123016

\bibitem{Gorini1976a}
Gorini V, Kossakowski A and Sudarshan E~C~G 1976 {\em J. Math. Phys.\/} {\bf
  17} 821

\bibitem{Lindblad1976}
Lindblad G 1976 {\em Commun. Math. Phys.\/} {\bf 48} 119

\bibitem{Chruscinski2017a}
Chru{\'{s}}ci{\'{n}}ski D and Pascazio S 2017 {\em Open Syst. Inf. Dyn.\/} {\bf
  24} 1740001

\bibitem{Galve2017}
Galve F, Mandarino A, Paris M~G~A, Benedetti C and Zambrini R 2017 {\em Sci.
  Rep.\/} {\bf 7} 42050

\bibitem{bellomo2017}
Bellomo B, Giorgi G~L, Palma G~M and Zambrini R 2017 {\em Phys. Rev. A\/} {\bf
  95} 043807

\bibitem{Hartmann2019}
Hartmann R and Strunz W~T 2019 {\em preprint arXiv:1906.02583\/}

\bibitem{gross1982superradiance}
Gross M and Haroche S 1982 {\em Phys. Rep.\/} {\bf 93} 301--396

\bibitem{ficek1987quantum}
Ficek Z, Tana{\'s} R and Kielich S 1987 {\em Physica A\/} {\bf 146} 452--482

\bibitem{walls1970higher}
Walls D~F 1970 {\em Z. Phys. A Hadr. Nucl.\/} {\bf 234} 231--241

\bibitem{schwendimann1972interference}
Schwendimann P 1972 {\em Z. Phys. A Hadr. Nucl.\/} {\bf 251} 244--253

\bibitem{Carmichael_1973}
Carmichael H~J and Walls D~F 1973 {\em J. Phys. A Math. Gen.\/} {\bf 6}
  1552--1564

\bibitem{Cresser1992}
Cresser J 1992 {\em J. Mod. Opt.\/} {\bf 39} 2187--2192

\bibitem{Scala2007}
Scala M, Militello B, Messina A, Piilo J and Maniscalco S 2007 {\em Phys. Rev.
  A\/} {\bf 75} 013811

\bibitem{Scala2007a}
Scala M, Militello B, Messina A, Maniscalco S, Piilo J and Suominen K~A 2007
  {\em J. Phys. A Math. Theor.\/} {\bf 40} 14527--14536

\bibitem{Migliore2011}
Migliore R, Scala M, Napoli A, Yuasa K, Nakazato H and Messina A 2011 {\em J.
  Phys. B At. Mol. Opt. Phys.\/} {\bf 44} 075503

\bibitem{Zoubi2003}
Zoubi H, Orenstien M and Ron A 2003 {\em Phys. Rev. A\/} {\bf 67} 063813

\bibitem{Saito2000}
Saito K, Takesue S and Miyashita S 2000 {\em Phys. Rev. E\/} {\bf 61}
  2397--2409

\bibitem{Henrich2005a}
Henrich M~J, Michel M, Hartmann M, Mahler G and Gemmer J 2005 {\em Phys. Rev.
  E\/} {\bf 72} 026104

\bibitem{Wichterich2007}
Wichterich H, Henrich M~J, Breuer H~P, Gemmer J and Michel M 2007 {\em Phys.
  Rev. E\/} {\bf 76} 031115

\bibitem{Capek2002}
{\v{C}}{\'{a}}pek V 2002 {\em Eur. Phys. J. B - Condens. Matter\/} {\bf 25}
  101--113

\bibitem{Novotny2002}
Novotn{\'{y}} T 2002 {\em Europhys. Lett.\/} {\bf 59} 648--654

\bibitem{DeChiara2018a}
{De Chiara} G, Landi G~T, Hewgill A, Reid B, Ferraro A, Roncaglia A~J and
  Antezza M 2018 {\em New J. Phys.\/} {\bf 20} 113024

\bibitem{Barra2015}
Barra F 2015 {\em Sci. Rep.\/} {\bf 5} 14873

\bibitem{PhysRevX.7.021003}
Strasberg P, Schaller G, Brandes T and Esposito M 2017 {\em Phys. Rev. X\/}
  {\bf 7} 021003

\bibitem{PhysRevA.90.063815}
Joshi C, \"Ohberg P, Cresser J~D and Andersson E 2014 {\em Phys. Rev. A\/} {\bf
  90} 063815

\bibitem{Manrique2015}
Manrique P~D, Rodr{\'{i}}guez F, Quiroga L and Johnson N~F 2015 {\em Adv.
  Condens. Matter Phys.\/} {\bf 2015} 1--7

\bibitem{Purkayastha2016}
Purkayastha A, Dhar A and Kulkarni M 2016 {\em Phys. Rev. A\/} {\bf 93} 062114

\bibitem{Santos2016}
Santos J~P and Landi G~T 2016 {\em Phys. Rev. E\/} {\bf 94} 062143

\bibitem{Decordi2017}
De{\c{c}}ordi G and Vidiella-Barranco A 2017 {\em Opt. Commun.\/} {\bf 387}
  366--376

\bibitem{Stockburger2017}
Stockburger J~T and Motz T 2017 {\em Fortschritte der Phys.\/} {\bf 65} 1600067

\bibitem{Hewgill2018}
Hewgill A, Ferraro A and {De Chiara} G 2018 {\em Phys. Rev. A\/} {\bf 98}
  042102

\bibitem{PhysRevA.98.052123}
Naseem M~T, Xuereb A and M\"ustecapl\ifmmode \imath \else \i
  \fi{}o\ifmmode~\breve{g}\else \u{g}\fi{}lu O~E 2018 {\em Phys. Rev. A\/} {\bf
  98} 052123

\bibitem{seah}
Seah S, Nimmrichter S and Scarani V 2018 {\em Phys. Rev. E\/} {\bf 98} 012131

\bibitem{raja2018thermodynamic}
Raja S~H, Borrelli M, Schmidt R, Pekola J~P and Maniscalco S 2018 {\em Phys.
  Rev. A\/} {\bf 97} 032133

\bibitem{PhysRevLett.114.220601}
Cui J, Cirac J~I and Ba\~nuls M~C 2015 {\em Phys. Rev. Lett.\/} {\bf 114}
  220601

\bibitem{Benatti2005}
Benatti F and Floreanini R 2005 {\em Int. J. Mod. Phys. B\/} {\bf 19}
  3063--3139

\bibitem{Dumcke1979}
D{\"{u}}mcke R and Spohn H 1979 {\em Z. Phys. B Con. Mat.\/} {\bf 34} 419--422

\bibitem{Suarez1992}
Su{\'{a}}rez A, Silbey R and Oppenheim I 1992 {\em J. Chem. Phys.\/} {\bf 97}
  5101--5107

\bibitem{Gaspard1999}
Gaspard P and Nagaoka M 1999 {\em J. Chem. Phys.\/} {\bf 111} 5668--5675

\bibitem{Wilkie2001}
Wilkie J 2001 {\em J. Chem. Phys.\/} {\bf 114} 7736--7745

\bibitem{Benatti2003}
Benatti F, Floreanini R and Piani M 2003 {\em Phys. Rev. A\/} {\bf 67} 042110

\bibitem{Ishizaki2009}
Ishizaki A and Fleming G~R 2009 {\em J. Chem. Phys.\/} {\bf 130} 234110

\bibitem{Argentieri2014}
Argentieri G, Benatti F, Floreanini R and Pezzutto M 2014 {\em Europhys.
  Lett.\/} {\bf 107} 50007

\bibitem{Jeske2015}
Jeske J, Ing D~J, Plenio M~B, Huelga S~F and Cole J~H 2015 {\em J. Chem.
  Phys.\/} {\bf 142} 064104

\bibitem{Sakurai}
Sakurai J~J 1994 {\em Modern Quantum Mechanics\/} (Addison-Wesley)

\bibitem{NielsenChuang}
Nielsen M~A and Chuang I 2010 {\em Quantum Computation and Quantum Information:
  10th Anniversary Edition\/} (Cambridge University Press)

\bibitem{aolita2015open}
Aolita L, De~Melo F and Davidovich L 2015 {\em Rep. Prog. Phys.\/} {\bf 78}
  042001

\bibitem{PhysRevA.81.012105}
Benatti F, Floreanini R and Marzolino U 2010 {\em Phys. Rev. A\/} {\bf 81}
  012105

\bibitem{PhysRevA.74.024304}
Ficek Z and Tana\ifmmode~\acute{s}\else \'{s}\fi{} R 2006 {\em Phys. Rev. A\/}
  {\bf 74} 024304

\bibitem{tanas2011sudden}
Tana{\'s} R 2011 Sudden death and sudden birth of entanglement {\em Quantum
  Dynamics And Information\/} (World Scientific) pp 179--198

\bibitem{galve2017quantum}
Galve F, Giorgi G~L and Zambrini R 2017 Quantum correlations and
  synchronization measures {\em Lectures on General Quantum Correlations and
  their Applications\/} (Springer) pp 393--420

\bibitem{Alicki1979}
Alicki R 1979 {\em J. Phys. A Math. Gen.\/} {\bf 12} L103--L107

\bibitem{PhysRevA.85.032110}
Salmilehto J, Solinas P and M\"ott\"onen M 2012 {\em Phys. Rev. A\/} {\bf 85}
  032110

\bibitem{giorgi2016}
Giorgi G~L, Galve F and Zambrini R 2016 {\em Phys. Rev. A\/} {\bf 94} 052121

\bibitem{Leggett1987}
Leggett A~J, Chakravarty S, Dorsey A~T, Fisher M~P~A, Garg A and Zwerger W 1987
  {\em Rev. Mod. Phys.\/} {\bf 59} 1--85

\bibitem{Majer2007a}
Majer J, Chow J~M, Gambetta J~M, Koch J, Johnson B~R, Schreier J~A, Frunzio L,
  Schuster D~I, Houck A~A, Wallraff A, Blais A, Devoret M~H, Girvin S~M and
  Schoelkopf R~J 2007 {\em Nature\/} {\bf 449} 443--447

\bibitem{Mastellone2008}
Mastellone A, D'Arrigo A, Paladino E and Falci G 2008 {\em Eur. Phys. J. Spec.
  Top.\/} {\bf 160} 291--300

\bibitem{Wendin2007a}
Wendin G and Shumeiko V~S 2007 {\em Low Temp. Phys.\/} {\bf 33} 724--744

\bibitem{Stokes2018}
Stokes A and Nazir A 2018 {\em New J. Phys.\/} {\bf 20} 043022

\end{thebibliography}

\end{document}